\documentclass[12pt,a4paper,twoside]{article}
\usepackage{times}
\usepackage[utf8]{inputenc}
\usepackage{url}
\usepackage{scalefnt}
\usepackage{fancyhdr}
\usepackage{amsthm}
\usepackage{amsmath}
\usepackage{amssymb}
\usepackage{amsfonts,amssymb,amsxtra}
\usepackage[mathcal]{euscript}
\usepackage{psfrag}
\usepackage{graphicx}
\usepackage{lipsum}
\usepackage{bbm}
\usepackage{color, colortbl}
\usepackage{tabularx}
\usepackage{longtable} 
\usepackage{epsfig}
\usepackage{float}
\usepackage{tikz}
\usetikzlibrary{shapes.geometric, arrows}
\usepackage{import}
\usepackage{mathptmx}
\usepackage[scaled=0.92]{helvet}
\usepackage{lastpage}
\usepackage{setspace}
\usepackage{geometry}
\usepackage{indentfirst}
\usepackage{color}
\usepackage{colortbl}
\usepackage{subcaption}
\usepackage{multicol}
\usepackage{multirow}
\usepackage{arydshln}
\usepackage{natbib}
\usepackage[titletoc]{appendix}
\usepackage{booktabs}
\usepackage{caption}
\usepackage{dsfont}
\usepackage{listings}
\usepackage{mathtools}
\usepackage{amsmath}
\usepackage{hyperref}
\usepackage{bbm}
\usepackage[figuresright]{rotating}
\usepackage{makecell}
\usepackage{nccmath}

\tikzstyle{arrow} = [thick,->,>=stealth]
\tikzstyle{startstop} = [rectangle, minimum width=8cm, minimum height=1cm,text centered, draw=black]
\definecolor{gray}{rgb}{0.9,0.9,0.9}
\definecolor{coral}{rgb}{1.0, 0.5, 0.31}
\definecolor{coral2}{rgb}{0.93,0.42,0.31}
\definecolor{cyan3}{rgb}{0,0.80,0.80}
\definecolor{darkred}{rgb}{0.55,0,0}
\definecolor{green4}{rgb}{0,0.55,0}
\definecolor{tan4}{rgb}{0.55,0.35,0.17}
\definecolor{slategrey}{rgb}{0.44,0.5,0.56}
\definecolor{cadetblue4}{rgb}{0.33,0.53,0.55}

\begin{document}

\begin{center}
\large{\textbf{\textbf{Joint model for zero-inflated data combining fishery-dependent and fishery-independent sources}}}
\end{center}
\vspace{0.2cm}

\begin{center}
Daniela Silva*\\
Division of Modeling and Management of Fishery Resources, Portuguese Institute for the Sea and Atmosphere (IPMA), Lisbon, Portugal,\\
\textit{daniela.dasilva@ipma.pt}\\
\vspace{0.3cm}

Raquel Menezes\\
Centre of Mathematics, Minho University, Guimarães, Portugal,\\
\textit{rmenezes@math.uminho.pt}\\
\vspace{0.3cm}

Gonçalo Araújo\\
Nova School of Business and Economics, Nova University Lisbon, Lisbon, Portugal,\\
CCMar - Centre of Marine Sciences, University of Algarve, Faro, Portugal,\\
University of Algarve, Faro, Portugal,\\
\textit{goncalo.araujo@novasbe.pt}\\
\vspace{0.3cm}

Renato Rosa\\
Centre of Business and Economics Research, Coimbra University, Coimbra, Portugal,\\
\textit{renato.rosa@novasbe.pt}\\
\vspace{0.3cm}

Ana Moreno\\
Division of Modeling and Management of Fishery Resources, Portuguese Institute for the Sea and Atmosphere (IPMA), Lisbon, Portugal,\\
\textit{amoreno@ipma.pt}\\
\vspace{0.3cm}

Alexandra Silva\\
Division of Modeling and Management of Fishery Resources, Portuguese Institute for the Sea and Atmosphere (IPMA), Lisbon, Portugal,\\
Marine and Environmental Sciences Centre (MARE), University of Évora, Évora, Portugal,\\
\textit{asilva@ipma.pt}\\
\vspace{0.3cm}

Susana Garrido\\
Division of Modeling and Management of Fishery Resources, Portuguese Institute for the Sea and Atmosphere (IPMA), Lisbon, Portugal,\\
\textit{susana.garrido@ipma.pt}
\end{center}
\vspace{0.2cm}

\begin{center}
\section*{Abstract}
\end{center}

Accurately identifying spatial patterns of species distribution is crucial for scientific insight and societal benefit, aiding our understanding of species fluctuations. The increasing quantity and quality of ecological datasets present heightened statistical challenges, complicating spatial species dynamics comprehension. Addressing the complex task of integrating multiple data sources to enhance spatial fish distribution understanding in marine ecology, this study introduces a pioneering five-layer Joint model.

The model adeptly integrates fishery-independent and fishery-dependent data, accommodating zero-inflated data and distinct sampling processes. A comprehensive simulation study evaluates the model performance across various preferential sampling scenarios and sample sizes, elucidating its advantages and challenges.

Our findings highlight the model's robustness in estimating preferential parameters, emphasizing differentiation between presence-absence and biomass observations. Evaluation of estimation of spatial covariance and prediction performance underscores the model's reliability. Augmenting sample sizes reduces parameter estimation variability, aligning with the principle that increased information enhances certainty.

Assessing the contribution of each data source reveals successful integration, providing a comprehensive representation of biomass patterns. Empirical validation within a real-world context further solidifies the model's efficacy in capturing species' spatial distribution. This research advances methodologies for integrating diverse datasets with different sampling natures further contributing to a more informed understanding of spatial dynamics of marine species.\\

\textbf{Keywords:} Species distribution model; Integrating data sources; Preferential sampling; Geostatiscal modeling; Fish data.\\

\section{Introduction}\label{sec:introd}

Scientific tools capable of identifying species distribution patterns are important, not only from a scientific point of view but also from a societal one, as they contribute to improving knowledge of causes of species fluctuations. In ecological field surveys, observations can be gathered at different spatial locations and time stamps. With the increasing quantity and quality of available datasets a higher complexity of statistical issues arises, which also means that understanding the spatio-temporal dynamics of species is becoming more challenging \citep{Minaya2018}.

The combination of multiple datasets has demonstrated significant advantages across various research fields \citep{Steele2008, Kirk2012, Ferreras2021}, and in particular within the realm of species distribution \citep{Doser2021, Tehrani2022}. This integration aims not only to leverage the wealth of information inherent in each dataset but also to potentiate and enhance the precision of predictive models, marking a critical step toward advancing our understanding of species distribution dynamics.

In the scope of fisheries science, there are two main types of data collection methods to assess the status of fish populations and make informed management decisions, fishery-independent data (FID) and fishery-dependent data (FDD). 

FID refers to information collected through methods that are independent of fishing activities. It involves conducting surveys or research specifically designed to assess fish populations, often using standardized sampling techniques. These surveys can be conducted using various tools such as fishing gears or acoustic technologies. FID provides a more reliable assessment of fish population size, abundance, and distribution, as it is not influenced by changes in fishing effort or selective harvesting practices. It helps estimate population parameters, such as growth rates, recruitment, and survival rates \citep{Ault1998}, which are crucial for determining the health and sustainability of fish stocks. FDD refers to information collected directly from fishing activities, such as commercial fishing. This data is typically collected through logbooks, fishery surveys, or monitoring programs. As a whole, it provides valuable insights into the characteristics of fish catches, such as size distribution, species composition, and catch rates. It also helps estimate fishing mortality rates and monitor changes in fishing effort over time \citep{Rosenberg2005}. This information is essential for managing fisheries, setting catch limits, and implementing conservation practices.

Research surveys are usually limited to specific operational time frames, occurring once or twice a year, covering a larger spatial region. Conversely, data from commercial vessels often has a higher frequency of sampling in time due to the nature of the activity, but subject to the preferential selection of the sampling locations. Hence, sampling designs for FDD and FID are tailored to their specific objectives. Nevertheless, both sources provide insights into fish populations dynamics and contribute to fisheries management. As a result, information obtained from the two data sources is different and can complement each other. However, joint modeling of these sources  requires approaches capable of dealing with different sampling designs, as classical tools handle standardized sampling designs but not the preferential nature of commercial data. Indeed, preferential sampling (PS) affects both the resulting predictive surface \citep{Diggle2010} and parameters estimation  \citep{Gelfand2012}.

Recent research on PS has increasingly referenced the model framework introduced by \citet{Diggle2010}, which has served as a point of departure for further studies exploring this phenomenon. This class of models, often articulated as a two-part model, posits that observed locations originate from a Poisson process with an intensity function linked to the underlying field. Simultaneously, the spatial process of interest is modeled under a suited distribution, encapsulating the sample design through the Log-Gaussian Cox process. \citet{Pati2011} extended this approach by introducing covariates and random effects into the model.

Moreover, species distribution data often implies residual spatial autocorrelation, which may arise due to the non-consideration of important environmental factors such as climate conditions that influence species distribution, as well as intrinsic factors such as competition, dispersal, and aggregation \citep{Guelat2018}.

Another significant challenge in ecological datasets, particularly in species distribution data, is the high prevalence of zero values. The incorporation of zero-inflated (ZI) approaches is pivotal in overcoming the challenges posed by excess zeros. As a result, ZI models \citep{Lambert1992} have become central to research in species distribution models (SDMs). These approaches are essential for achieving a more accurate representation of true absences \citep{MacKenzie2006}, thereby enhancing model fit and improving predictive accuracy \citep{Arroita2012}.

While FID and FDD have long been employed in studying fish distribution, their integration into a unified model for more informative outputs has only recently gained attention in the scientific literature. An example of such integration is evident in the work of \citet{Rufener2021}, where an integrated statistical model was proposed to infer fish abundance distribution by leveraging scientific survey and commercial fisheries data. \citet{Rufener2021} introduced a three-layered structure, comprising the latent process, the observation processes for survey and commercial data, and the associated parameters. The latent process delineates expected species numbers through a log-link linear function, incorporating a spatio-temporal structured random effect and a set of explanatory variables. Conditioned on the latent process, the observation processes incorporate catchability effects, varying based on the data source. For survey data, catchability is expressed through indicators of research vessels, while gear indicators are also considered for the catchability associated with commercial data.

In the work presented by \citet{Alglave2022}, a hierarchical SDM combined scientific survey and commercial catch data. This model comprises four layers, encompassing observations from both survey and commercial data, the sampling processes (one for each vessel), the latent field representing fish biomass relative density, and the associated parameters. The latent field is modeled using a log-link linear function with explanatory covariates and a spatially structured random effect. Sampling processes are conceptualized as Inhomogeneous Poisson Processes (IPPs), with the logarithm of the intensity function being a linear combination of an intercept, the logarithm of the relative biomass, and a spatially structured residual effect. The degree of PS is quantified by the scaling parameter between the relative biomass field and the sampling process. Finally, the observation process is captured through a ZI lognormal model conditional on relative biomass.

In this study, we present a novel spatial Joint model designed for ZI data, aiming to infer the spatial distribution of fish by integrating information from both FID and FDD, while effectively addressing the challenges associated with PS. Our proposal consists of a five layers model emphasizing the differences between modeling presence/absence observations and biomass observations. To comprehensively evaluate the performance, challenges, and advantages of our Joint model, we conducted a simulation study. This simulation encompassed various scenarios of PS and configurations of sample sizes for both simulated FDD and FID sources. Finally, the proposed model was applied to a real-world case study to estimate the spatial distribution of European sardine (\textit{Sardina pilchardus}, Walbaum 1792) along the southern coast of Portugal.

The work is structured into five sections. In Section 2, we detail the proposed Joint model, outline the simulation study, and describe the data underlying the case study. Section 3 presents the results obtained from both the simulation study and the application of our proposed model to a specific case study, offering an evaluation of the model and real-world insights into its practical utility. Section 4 provides a comprehensive discussion of the results, exploring their implications and potential avenues for future research.

\section{Material and Methods}\label{sec:mat_met}

\subsection{Joint model}
\label{ssec:jmodel}

To infer species distribution taking advantage of the information provided by both FID and FDD sources, we propose a joint hierarchical model with five layers: presence-absence observations, biomass observations under presence, the sampling process, the latent fields, and the parameters.

\subsubsection{Observations}

Let us denote the spatial biomass process by $\mathbf{S}=\{s_{\mathbf{x}_1},\cdots,s_{\mathbf{x}_n}\}$ at location $\mathbf{x}_i \in \mathcal{A} \subset \mathbb{R}^2$, where $\mathcal{A}$ is the study region and $n$ represents dimension of the data. The presence-absence process (PAP) $\mathbf{Z}=\{z_{\mathbf{x}_1},\cdots,z_{\mathbf{x}_n}\}$, with presence probability $\pi_{\mathbf{x}_i}$, takes the binary value 0 if no species was observed at location $\mathbf{x}_i$, and 1 otherwise. The biomass process under the presence $\mathbf{Y}=\mathbf{S} \vert (\mathbf{Z}=1)=\{y_{\mathbf{x}_1},\cdots,y_{\mathbf{x}_n}\}$ takes  the strictly positive values of the biomass process $\mathbf{S}$.

Consequently, the distribution of the process of interest $\mathbf{S}$ is given by the product of the distribution of the PAP $\mathbf{Z}$ and the distribution of the biomass process under the presence $\mathbf{Y}$ such that
\begin{equation}
P(S_i=s_i)=P(S_{\mathbf{x}_i}=s_{\mathbf{x}_i})= \left \{ \begin{matrix} 1-\pi_{\mathbf{x}_i}&~\text{if}~&s_{\mathbf{x}_i}=0\\ 
\pi_{\mathbf{x}_i}~p(s_{\mathbf{x}_i} \vert \mu_{\mathbf{x}_i})&~\text{if}~&s_{\mathbf{x}_i}>0
\end{matrix}\right.
\label{eq_theory_hurdle}
\end{equation}
where $p(s_{\mathbf{x}_i} \vert \mu_{\mathbf{x}_i})$ represents a probability mass function for $\mathbf{Y}$, the biomass process under presence (e.g., Gamma and Log-normal distributions). The same is observed for the main statistics of the interest process, mean $E[\mathbf{S}]=E[\mathbf{Z}]E[\mathbf{Y}]$ and median $F_{\mathbf{S}}(0.5)=E[\mathbf{Z}]F_{\mathbf{Y}}(0.5)$ \citep{Silva2023}.

We propose a two-part model (\eqref{eq:logit} and \eqref{eq:log}) designed for the inference of species biomass distribution. This model is specifically crafted to accommodate ZI data, taking into account the distinct conditions influencing both the PAP \eqref{eq:logit} and the biomass process under presence of the species \eqref{eq:log}. PAP $\mathbf{Z}$ is assumed to come from a Bernoulli distribution with probability $\pi$ such that $Z_{\mathbf{x}_i} \sim Bernoulli(\pi_{\mathbf{x}_i})$. The biomass process under the presence $\mathbf{Y}$ requires a continuous distribution such as Gamma distribution with shape parameter $a_{\mathbf{x}_i}=\mu_{\mathbf{x}_i}^2/\upsilon^2$ and scale parameter $b_{\mathbf{x}_i}=\upsilon^2/\mu_{\mathbf{x}_i}$, that is, $Y_{\mathbf{x}_i} \sim Gamma(a_{\mathbf{x}_i},b_{\mathbf{x}_i})$. $\mu_{\mathbf{x}_i}$ and $\upsilon$ represent the mean for location $\mathbf{x}_i$ and the standard deviation of biomass under presence, respectively.
\begin{align}
    logit(\pi_{\mathbf{x}_i})&=\alpha' + V_{\mathbf{x}_i}\label{eq:logit}\\
    log(\mu_{\mathbf{x}_i}) &= \alpha + U_{\mathbf{x}_i}\label{eq:log}.
\end{align}

$\alpha$ and $\alpha'$ parameters denote the intercepts of the linear predictors for the corresponding process. $\mathbf{U}_{\mathbf{X}}$ and $\mathbf{V}_{\mathbf{X}}$ are spatial latent fields as described below.

\subsubsection{Latent fields}
\label{sssec:latent}
The biomass process is modeled through two spatial fields $\mathbf{U}_{\mathbf{X}}$ and $\mathbf{V}_{\mathbf{X}}$, representing the spatial structure associated to $\mathbf{Y}$ and $\mathbf{Z}$, respectively. Each latent field denotes the spatial dependency and variation that is accounted for through a zero-mean Gaussian Markov Random Field (GMRF) with a Mat\'{e}rn covariance function $M(\mathbf{x},\mathbf{x'};\phi,\sigma, \nu)$ with spatial range $\phi$, marginal variance $\sigma^2$ and smoothing parameter $\nu$ such that $\mathbf{U}_{\mathbf{X}},\mathbf{V}_{\mathbf{X}} \sim GMRF(\mathbf{0},M(\mathbf{x},\mathbf{x'};\phi,\sigma,\nu))$.

\subsubsection{Sampling process}

Let us denote the spatial point processes underlying FID and FDD by $\mathbf{X}^{S}$ and $\mathbf{X}^{C}$, respectively\footnote{$S$ is the acronym for Survey in $\mathbf{X}^{S}$ and $C$ for commercial in $\mathbf{X}^{C}$.}. The intensity of a point process can exhibit either spatial constancy, yielding a homogeneous or stationary pattern, or spatial variability with a discernible spatial trend, resulting in an inhomogeneous pattern. Hence, the assumption of stationarity may prove unrealistic in certain applications. This is particularly evident when the process of interest dictates the data locations $\mathbf{x}^{C}$, and there exists stochastic dependence between this process and the one under consideration, as is often the case in FDD. Conversely, FID offers a more impartial representation of fish abundance distribution (e.g., random scheme and  systematic sampling). In this scenario, the sampling locations $\mathbf{x}^{S}$ are chosen independently of the process of interest, providing a democratic sample that avoids the biases introduced by stochastic dependencies.

To generalize the definition of the spatial point process $\mathbf{X}^S$ and account for various potential sampling schemes, it is modeled as a Homogeneous Poisson Process (HPP) with a constant mean $\lambda^{HPP}$. Thus, $\mathbf{X}^{S}$ is an independent process such that $\mathbf{X}^{S} \sim HPP(\lambda^{HPP})$.

Following \cite{Diggle2010}, the set of fishing locations $\mathbf{X}^C$ is modeled conditionally on $\mathbf{U}_{\mathbf{X}}$ and $\mathbf{V}_{\mathbf{X}}$ as an IPP with intensity function $\lambda_{\mathbf{x}_i^C}$, $\mathbf{X}^C \sim IPP(\lambda_{\mathbf{x}^C})$. The logarithm of the intensity function is expressed as:
\begin{align}
    log(\lambda_{\mathbf{x}_i^C})=\alpha''+ &\beta' V_{\mathbf{x}_i^C} + \beta U_{\mathbf{x}_i^C}. \label{eq:ipp}
\end{align}
Therefore, the intensity function of the IPP is described by the logarithmic link function of the linear combination of the intercept $\alpha''$ and the latent effects $U_{\mathbf{x}_i^C}$ and $V_{\mathbf{x}_i^C}$.
$\beta'$ and $\beta$ quantify the degree of PS by scaling the relationship between the local fishing intensity and the local value of each process of interest $\mathbf{Z}$ and $\mathbf{Y}$, respectively.

\begin{figure}
    \centering
    \includegraphics[width=0.65\textwidth]{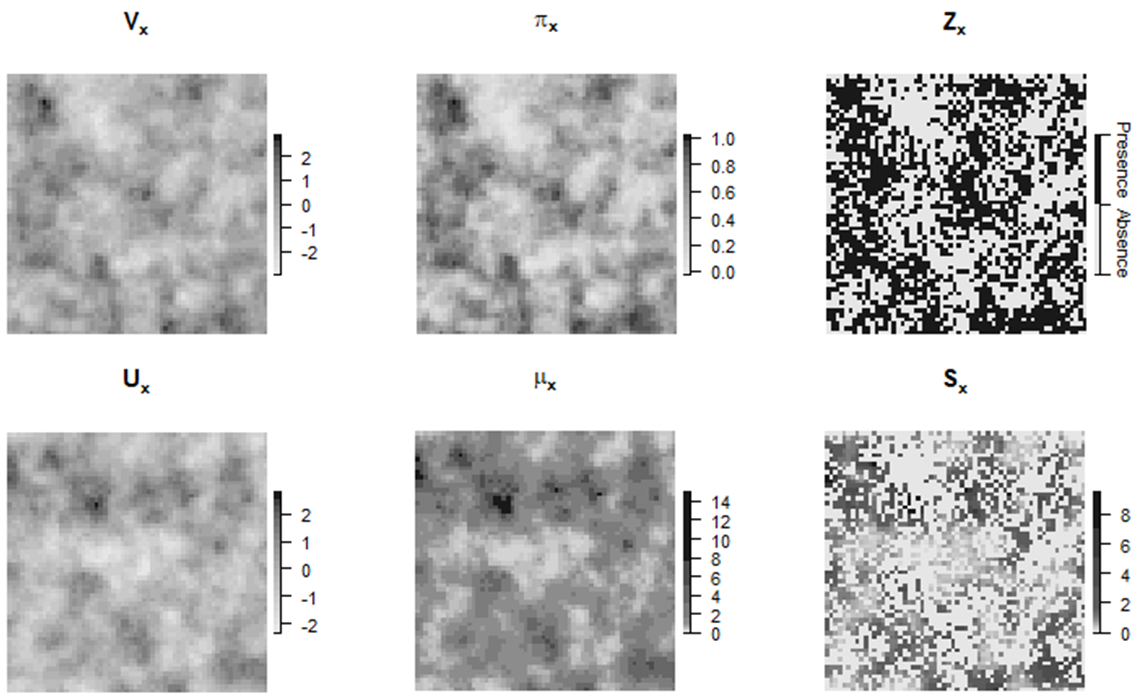}
    \caption{Example of simulated latent fields. $\mathbf{V}_{\mathbf{X}}$ and $\mathbf{U}_{\mathbf{X}}$ represent realizations of the simulated GMRFs. $\pi_{\mathbf{X}}$ and $\mu_{\mathbf{X}}$ are the probability field of species presence and mean field of species biomass under presence given $\mathbf{V}_{\mathbf{X}}$ and $\mathbf{U}_{\mathbf{X}}$, respectively. $\mathbf{Z}_{\mathbf{X}}$ represents realizations of the PAP determined by $\pi_{\mathbf{X}}$. The biomass process $\mathbf{S}_{\mathbf{X}}$ is derived as $\mathbf{S}_{\mathbf{X}}=\mathbf{Z}_{\mathbf{X}}\cdot\mathbf{Y}_{\mathbf{X}}$, where $\mathbf{Y}_{\mathbf{X}}$ represents realizations of the biomass process under presence determined by $\mu_{\mathbf{X}}$.}
    \label{fig:simu_data}
\end{figure}

Figure \ref{fig:simu_data} illustrates a realization of each process contributing to the determination of a single realization of the process $\mathbf{S}$.

\subsection{Inference and estimation}
\label{ssec:inf_est}

The model estimation and parameter inference are conducted through Laplace Approximation \citep{Skaug2006}, whose joint distribution is presented in Appendix \ref{sec:theory}. The model is formulated in \texttt{C++} and then fitted by using $\texttt{TMB R}$ package \citep{Kristensen2016}.

To derive each spatial latent field, we employ an approximation method based on stochastic partial differential equations (SPDEs), as introduced by \cite{Lindgren2011}. SPDE approach enables the approximation of a spatial continuous field, represented by a Mat\'{e}rn covariance function, to a GMRF which is discretly indexed. The adoption of this approximation is motivated by its computational efficiency. 

Parameterization of the latent fields is performed in terms of marginal variance $\sigma$ and range of influence $\phi$, enhancing the model interpretability and computational advantages. Under certain circumstances, a reparametrization of these parameters (as defined in Section \ref{sssec:latent}) proves to be advantageous. In this study, mainly dictated by the functionality of the $\texttt{TMB~R}$ package, our proposed model is implemented using the parameters $\kappa$ and $\tau$. Subsequently, assuming the fixed value of $\nu=1$, a reparametrization is undertaken to enhance the interpretability of the results according to $\phi=\frac{\sqrt{8\nu}}{\kappa}$ and $\sigma=\frac{\sqrt{\Gamma(\nu)}}{\sqrt{\Gamma(\nu+1)} \cdot \kappa^{\nu} \cdot \tau \cdot \sqrt{4\pi}}$.

\subsection{Scenarios of sampling}

Various combinations of the parameters $\beta'$ and $\beta$ give rise to distinct intensity functions of the point process, and consequently to diverse sampling scenarios. These scenarios may range from extremes, where sampling is solely contingent on either interest process $\mathbf{Z}$ or $\mathbf{Y}$, to situations where it is dependent on both processes. This array of scenarios enables the projection of real-world situations in fishery science, as fishermen often concentrate their efforts on sampling based on their prior knowledge of the species. For instance, fishermen seek regions with higher species abundance and simultaneously direct their efforts toward locations where the species is present. Another scenario arises when fishermen exclusively target regions where the species is present, without specific concern for the quantity captured. This may occur due to established fishing quotas and restrictions.

\begin{figure}
    \centering
    \includegraphics[width=0.65\textwidth]{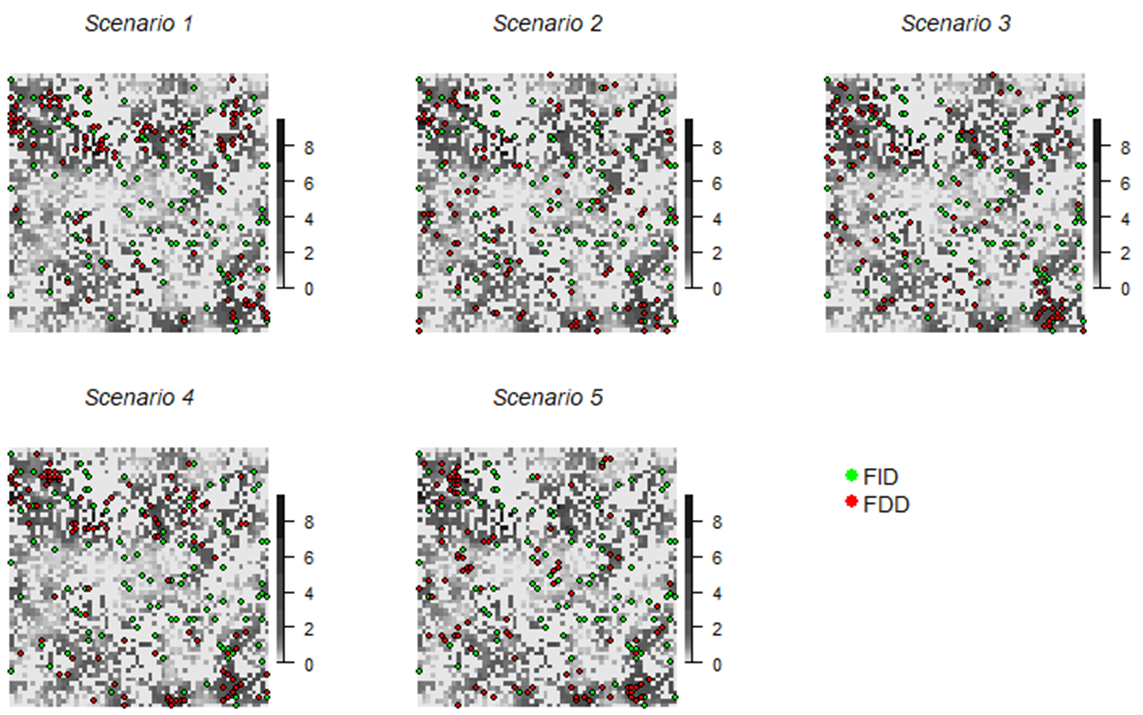}
    \caption{Examples of simulated both FID and FDD locations across sampling scenarios (\textit{Scenario 1}: $\beta'=0$ and $\beta=2$; \textit{Scenario 2}: $\beta'=1$ and $\beta=0.5$; \textit{Scenario 3}: $\beta'=1$ and $\beta=1$; \textit{Scenario 4}: $\beta'=1$ and $\beta=2$; \textit{Scenario 5}: $\beta'=2$ and $\beta=0$). \textcolor{green}{Green} points represent sample locations for simulated FID (Fishery-independent data), and \textcolor{red}{red} points identify the sample locations for simulated FDD (Fishery-dependent data). The depicted latent field is $\mathbf{S}$, and each data source has a dimension of 100.}
    \label{fig:scenarios}
\end{figure}

Below, we enumerate some of these scenarios whose representation of the set of sample locations is available in Figure \ref{fig:scenarios}.
\begin{itemize}
    \item \textit{Scenario 1:} Strong PS dependent on $\mathbf{Y}$\\
The sampling process for simulated FDD is entirely and strongly contingent on the biomass process under presence. Hence, $\beta' = 0$ and $\beta = 2$.
    \item \textit{Scenario 2:} Moderate PS dependent on $\mathbf{Z}$ and weak PS dependent on $\mathbf{Y}$\\
The sampling process depends on both the PAP and biomass under the presence process, with parameters set at $\beta' = 1$ and $\beta = 0.5$.
    \item \textit{Scenario 3:} Moderate PS dependent on $\mathbf{Z}$ and $\mathbf{Y}$\\
The FDD was simulated under the combination $\beta' = 1$ and $\beta = 1$.
    \item \textit{Scenario 4:} Moderate PS dependent on $\mathbf{Z}$ and strong PS dependent on $\mathbf{Y}$\\
The sampling process for FDD is dependent on both processes of interest, with a higher weight assigned to the biomass process under presence $\mathbf{Y}$. In this setting, $\beta' = 1$ and $\beta = 2$.
    \item \textit{Scenario 5:} Strong PS dependent on $\mathbf{Z}$\\
The sampling locations for simulated FDD are contingent on the PAP of the species. The preferentiality parameters are set such that $\beta'=2$ and $\beta = 0$.
\end{itemize}

\subsection{Simulation-estimation experiments}

Simulation-estimation experiments are carried out to evaluate the performance of the proposed model across various data and model configurations. Each scenario is simulated on a regular $60 \times 60$ grid within the domain $[0,1] \times [0,1]$. Range and marginal variance  parameters are individually set for each GMRF, denoted as $\mathbf{U}_{\mathbf{X}}$ and $\mathbf{V}_{\mathbf{X}}$, to assess the model performance concerning distinct spatial dependencies of both responses $\mathbf{Z}$ and $\mathbf{Y}$. That is, assuming that both responses are governed by different processes. Specifically, $(\phi_{\mathbf{V}},\sigma_{\mathbf{V}}^2)=c(0.15,0.80)$ and $(\phi_{\mathbf{U}},\sigma_{\mathbf{U}}^2)=c(0.20,1.00)$. Additionally, the intercept parameters, namely $\alpha$, $\alpha'$, and $\alpha''$, were assumed to be zero across all scenarios.

To assess how the sample sizes of both FID ($n^S$) and FDD ($n^C$) influence the relative contribution of each data source, simulations are conducted with various combinations of sample sizes $C_n(n^S, n^C)$. These combinations are chosen to represent possible real-world situations, allowing the dimensions of both data sources to be equal or different.

The selected combinations include $C_n(100,100)$, representing a scenario where both data sources share identical dimensions. Recognizing that in practical situations, FDD often exhibits larger dimensions compared to FID due to factors such as financial constraints and time-intensive surveying, additional combinations are explored. These include $C_n(100,200)$ and a more asymmetric scenario $C_n(100,500)$. Conversely, to account for scenarios where a greater emphasis on FID may arise due to fishery restrictions or limited interest in specific species by fishermen, a combination with larger FID dimensions is considered, denoted as $C_n(200,100)$. This scenario reflects a less common, yet plausible.

In summary, the selected combinations provide a detailed exploration of the interplay between sample sizes of FID and FDD, capturing realistic scenarios ranging from balanced dimensions to instances where one data source dominates due to practical constraints and ecological considerations.

Furthermore, we conduct a comprehensive assessment of the contribution of each data source in \textit{Scenario 4} ($\beta'=1$ and $\beta=2$) through a comparative analysis of three models: the FDD model, the FID model, and the Joint model. The FDD model encompasses our proposed model but exclusively utilizes FDD data. In contrast, the FID model represents a simplified version of our proposed model, neglecting the modeling of the sampling process and relying solely on FID data. The selection of \textit{Scenario 4} ($\beta'=1$ and $\beta=2$) for this evaluation was deliberate, aiming to capture a scenario that closely mirrors real-world conditions and thus provides insights into the distinctive contributions of each data source.

To ensure robustness, each scenario and configuration is repeated 100 times, allowing for the capture of variability among replicates. 

\subsection{Performance metrics}

The assessment of the estimation performance of the proposed model involves a comprehensive analysis of various model outputs. The evaluation encompasses all estimated parameters, including the intercept, preferential, and spatial covariance parameters, as well as the spatial predictions.

To gauge the accuracy of intercept parameters estimation, the distribution of estimates is scrutinized across 100 replicas. Given the potential for asymmetric distributions in spatial covariance parameters across replicas, their estimation quality is performed through the identification of the median and the interquartile interval.

In addition to assessing parameters estimation, the predictive performance of the proposed method is thoroughly evaluated using three distinct metrics: RMSE, MAE, and the Hellinger distance \citep{Lecam1986}. These metrics provide a robust evaluation of the model's ability to generate spatial predictions that align closely with observed data. In this context, the Hellinger distance measures the similarity between the observed data and the predicted data distributions, ranging from 0, indicating equality, and 1, indicating ``total difference''.

\subsection{Case study}

Given the socioeconomic significance of the European sardine for Portugal and Spain, and the abundance of available data pertaining to this species, we undertake the task of predicting its spatial distribution within the Portuguese shelf. For illustrative purposes, our study focuses specifically on the southern region of the Portuguese coast, an important area for sardine fisheries. In our predictive modeling approach, we incorporate two data sources - FID and FDD - to comprehensively represent the two main sources of fishery data.

\subsubsection{Fishery-independent data}

\begin{figure}
    \centering
    \includegraphics[scale=0.5]{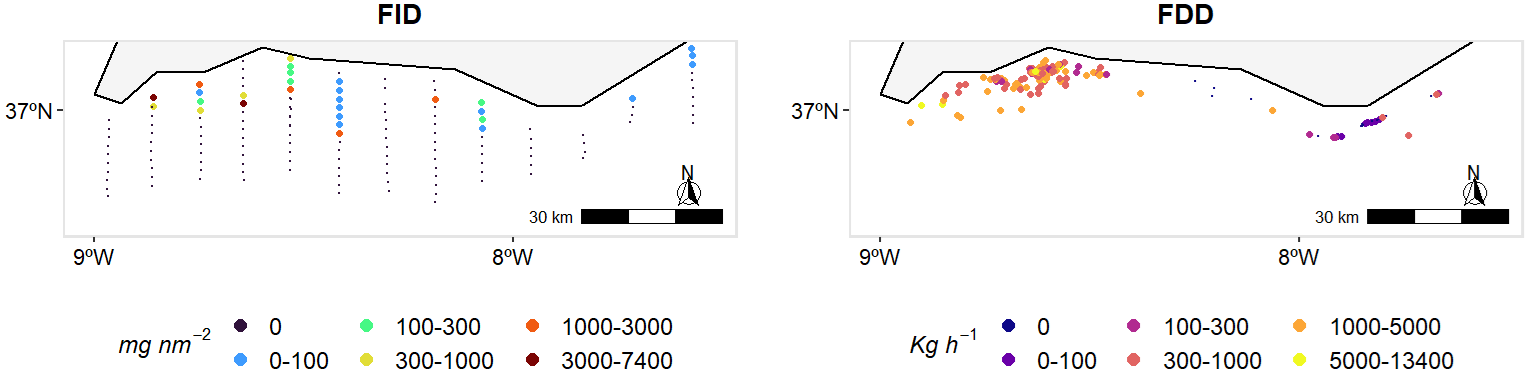}
   \caption{Observed biomass index of sardine off the southern coast of Portugal during 2017 from FDD source in $mg^2~nm^{-2}$ (first column) and from FDD source in $Kg~h^{-1}$ (second column).}
    \label{fig:observ_biomass}
\end{figure}

The spatial distribution of sardine biomass is assessed using data from the Portuguese spring acoustic (PELAGO) series (first row of Figure \ref{fig:observ_biomass}), conducted by the Portuguese Institute for the Sea and Atmosphere (IPMA) in Continental Portuguese waters during 2017.

\begin{table}
\centering
\caption{Summary of recorded locations of sardine off the southern coast of Portugal, along with the count and percentage of locations exhibiting strictly positive values (i.e., presences), for both data sources FID (PELAGO survey series data) and FDD (commercial data obtained through the AIS) during 2017. The table includes the overall estimate
of the nautical area-scattering coefficient (NASC) derived from FID and the total Catch Per Unit Effort (CPUE) from FDD for sardine.}
\begin{tabular}{cccc}
\toprule
\multicolumn{1}{c}{Data source} & \begin{tabular}[c]{@{}c@{}}Number of\\ locations\end{tabular} & \begin{tabular}[c]{@{}c@{}}Sardine positive\\ locations\end{tabular} & \begin{tabular}[c]{@{}c@{}}Total \\ estimated/captured\end{tabular} \\ \midrule
FID                       &         144                                                   &     29 (20\%)                                                        &     26142.06  $mg~nm^{-2}$                                                     \\
FDD                       &        151                                                       &         127 (84\%)                                                             &     182434 $Kg~h^{-1}$                                                        \\ \bottomrule
\end{tabular}
\label{tab:data}
\end{table}

The primary objective of the PELAGO series is to monitor the spatial distribution of abundance, biomass, and various biological parameters of sardines and other small pelagic fish. The survey design involves continuous daytime acoustic measurements along parallel transects, facilitated by a calibrated 38-kHz echosounder. Data processing includes integrating and averaging the resulting backscatter from the water column over 1 $nm$ intervals, expressed as nautical area-scattering coefficients [NASC; $S_A$ (in $m^2~nm^{-2}$)]. The inter-transect distance is consistently 6 $nm$. The detailed methodology underpinning the PELAGO series is outlined in \cite{Doray2021}.

Each NASC value, representing as a proportion of fish density, is utilized as a biomass proxy for each pair of coordinates (longitude and latitude). The FID source incorporates 144 sardine NASC values recorded in 2017, where the majority (about 80\%) represent species absence (Table \ref{tab:data}).

\subsubsection{Fishery-dependent data}

For the same area of interest, the FDD source consists of recent output data by \cite{Araujo2023} generated from Automatic Identification System (AIS) data obtained under a commercial licence for the Portuguese mainland purse seiners. Importantly, this commercial data aligns with the period when the scientific survey was conducted, ensuring consistency in the temporal scope to avoid variations in species distribution patterns that may occur throughout the year, specifically between April 24th and June 6th of 2017. The dataset from commercial source is standardized by fishing effort (total duration in hours of the fishing event), quantified in kilograms per hour ($Kg~h^{-1}$), enhancing comparability across samples. The FDD dataset comprises 151 commercial samples, providing valuable insights into the spatial distribution of sardine biomass in the studied region (second row of Figure \ref{fig:observ_biomass}). Conversely, the majority (about 84\%) of the FDD observations indicate species presence (Table \ref{tab:data}).

\subsubsection{Catchability effect}

Given the distinct biomass indices for sardines derived from the FID (measured in NASC units) and the FDD (expressed as catch in $Kg~h^{-1}$), the proposed model is applied to estimate the relative biomass index, denoted as the underlying process of interest $\mathbf{S}$. In this framework, we assume that the expected biomass index value, $\zeta_{\mathbf{x}j}$, for each vessel $j$ (whether associated with FID or FDD) and spatial location $\mathbf{x}$, is a function of the expected relative biomass $\mu_{\mathbf{x}}$ and a catchability parameter $k_j$, defined as
\begin{equation}
\zeta_{\mathbf{x}j}=k_j \times \mu_{\mathbf{x}}
\label{eq:relat_biom}
\end{equation}

The catchability parameter allows to adjust for measurement differences between the two data sources, to capture the vessel-specific differences in catch efficiency, and to ensure that the relative biomass estimates are comparable across sources allowing each index to be proportional to the underlying biomass $\mu_{\mathbf{x}}$.

In the present case study, the survey data were collected using a single research vessel, whereas the commercial data were obtained from observations across fifteen distinct fishing vessels. Accordingly, the index $j$ takes values $j=\,\cdots,16$ where $j=1$ corresponds to the survey vessel, and $j=2,\cdots,16$ represent the fifteen commercial vessels.

\section{Results}

\subsection{Simulation study}

\subsubsection{Evaluation of the estimation of preferential parameters}

\begin{figure}
    \centering
    \begin{subfigure}{0.6\linewidth}
        \includegraphics[width=\linewidth]{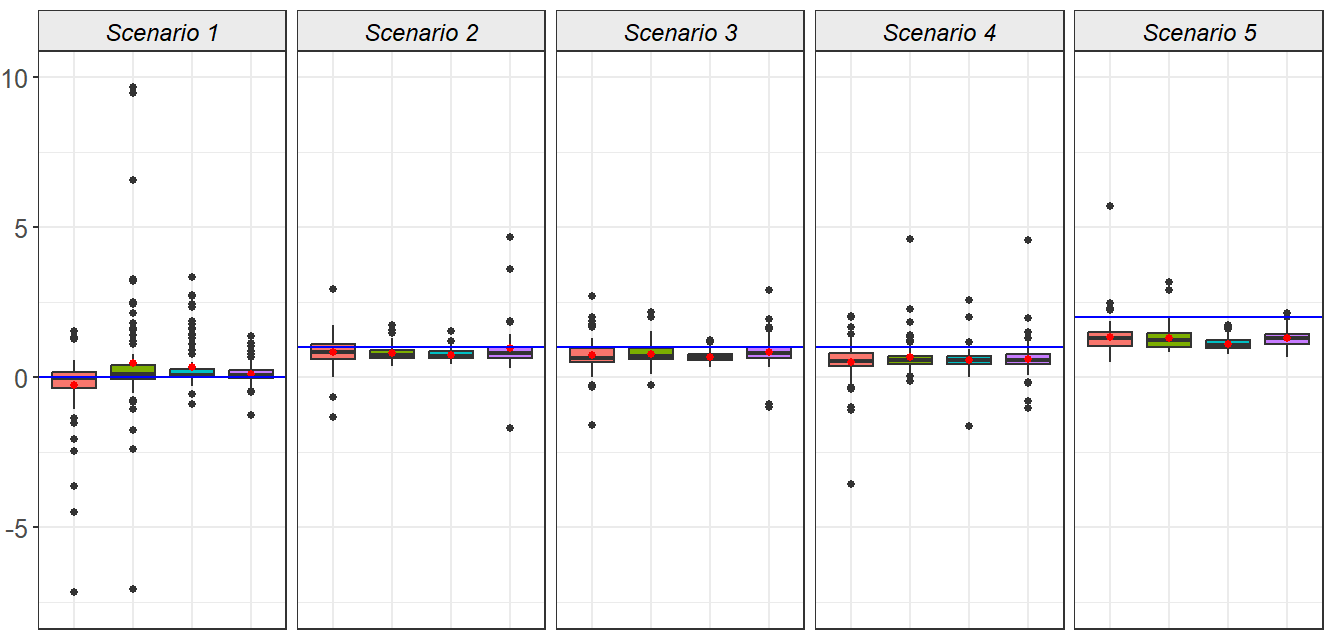}
        \caption{$\beta'$}
        \label{fig:beta_z}
    \end{subfigure}
    \begin{subfigure}{0.6\linewidth}
        \includegraphics[width=\linewidth]{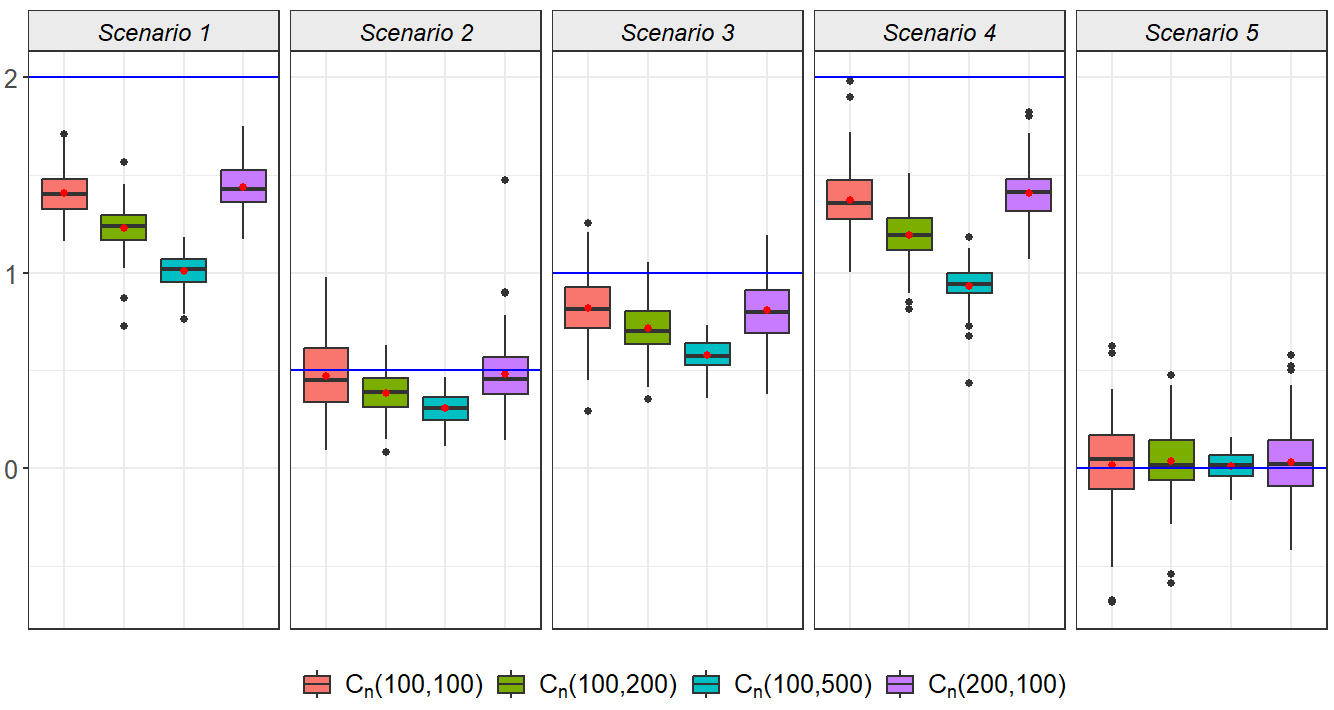}
        \caption{$\beta$}
        \label{fig:beta_y}
    \end{subfigure}
    \caption{Estimates of preferential parameters ($\beta'$ and $\beta$) across sampling scenarios (\textit{Scenario 1}: $\beta'=0$ and $\beta=2$; \textit{Scenario 2}: $\beta'=1$ and $\beta=0.5$; \textit{Scenario 3}: $\beta'=1$ and $\beta=1$; \textit{Scenario 4}: $\beta'=1$ and $\beta=2$; \textit{Scenario 5}: $\beta'=2$ and $\beta=0$) and combination of samples' dimensions $C_n(n^S, n^C)$. The \textcolor{red}{red} points represent the mean values of all 100 replicas and \textcolor{blue}{blue} lines identify the true values  assumed for the preferential parameters.}
    \label{fig:beta_by_comb_dim}
\end{figure}

The proposed model demonstrates a capacity to yield valuable estimates for preferential parameters (Figure \ref{fig:beta_by_comb_dim}). When assuming $\beta'$ or $\beta$ as zero, the model provides accurate estimates for these parameters. Moreover, under scenarios of moderate or weak effects of PS ($\beta'=1$ and/or $\beta=\{0.5,1\}$), the model reliably estimates preferential parameters for FDD dimension up to 200, albeit slight underestimation is observed for FDD dimensions of 500. In instances of a strong PS effect, while there is a tendency for parameter underestimation, the estimates remain statistically significant. 

Furthermore, increasing the dimensions of the FDD enables the selection of locations with lower corresponding values of the intensity function, thereby capturing samples with reduced values (see Appendix \ref{app:B}). This expansion results in a decreased mean for the process of interest. The augmentation of FDD sample size also emerges as a pivotal factor in reducing variability in parameter estimation.

Conversely, the increase in sample size of FID does not exert a discernible impact on the estimation of preferentiality parameters. This outcome aligns with expectations, as $\beta$ and $\beta'$ parameters are only utilized to describe the spatial arrangement of FDD. The lack of influence from FID on preferentiality parameter estimation underscores the distinct roles played by FID and FDD in shaping the precision and reliability of the parameter estimates.

\subsubsection{Evaluation of the estimation of spatial covariance parameters}

In most considered model configurations, encompassing various combinations of sample sizes and scenarios, we observe a pattern of slight overestimation in the range parameter, denoted as $\phi_{\mathbf{V}}$ (refer to Table \ref{tab:res_sp_param}). Generally, as sample sizes increase, the estimated $\phi_{\mathbf{V}}$ distantiates to its true value. \textit{Scenario 1}, characterized by a robust stochastic dependence influenced by the biomass process, yields more accurate $\phi_{\mathbf{V}}$ estimates irrespective of sample sizes. Conversely, \textit{Scenario 5}, representing a scenario where the PS is only influenced by the PAP, produces more biased estimates of the range parameter within the GMRF $\mathbf{V}_{\mathbf{X}}$. The precision of $\phi_{\mathbf{V}}$ estimation is found to be sensitive to the sample sizes, with increased variability as both sample sizes grow.

In all scenarios and across various combinations of sample sizes, accurate estimation of the $\sigma_{\mathbf{V}}$ parameter is observed. That is, the interquartile intervals consistently encompass the true value, affirming the method adequacy for estimating this parameter. Additionally, as the sample size of the FDD increases, there is a decrease in bias. Upon comparing all scenarios, it is evident that \textit{Scenario 2}, under the moderate dependence of the sampling process on the PAP and weak dependence on the biomass process, stands out by providing more accurate estimates for $\sigma_{\mathbf{V}}$.

Contrasting with the findings concerning the estimation of $\phi_{\mathbf{V}}$, the estimation of the range parameter $\phi_{\mathbf{U}}$ is reliable across all model configurations. Moreover, the augmentation of sample sizes generally induces higher estimates of the range parameter for biomass under the presence process. \textit{Scenario 1}, characterized by the absence of PS dependent on PAP, yields more biased estimates for the $\phi_{\mathbf{U}}$ parameter. This stands in contrast to \textit{Scenarios 3} and \textit{5} which result in more accurate estimates.

A prevalent accurate estimation of the $\sigma_{\mathbf{U}}$ parameter is observed for all of model configurations, despite the approximation to the true value with the augmentation of data dimension. 

In summary, the assessment of spatial covariance estimation reveals consistent patterns, including the tendency for accurate estimation of both range marginal variance parameters across various model configurations, except the observed overrestimation of $\phi_{\mathbf{V}}$. Additionally, the reliability of parameter estimation improves when the sample size of FID increases and the effect of PS on PAP is null or moderate.

\begin{table}
\caption{Median values (and interquartile intervals) for $\phi_{\mathbf{V}}=0.15$, $\sigma_{\mathbf{V}}=\sqrt{0.80}$, $\phi_{\mathbf{U}}=0.20$ and $\sigma_{\mathbf{U}}=1.00$ across sampling scenarios (\textit{Scenario 1}: $\beta'=0$ and $\beta=2$; \textit{Scenario 2}: $\beta'=1$ and $\beta=0.5$; \textit{Scenario 3}: $\beta'=1$ and $\beta=1$; \textit{Scenario 4}: $\beta'=1$ and $\beta=2$; \textit{Scenario 5}: $\beta'=2$ and $\beta=0$) and combinations of samples' dimensions $C_n(n^S, n^C)$.}
\centering
\footnotesize{
\begin{tabular}{ccrrrr}
\hline
Param. & \textit{Scen.}  & $C_n(100,100)$ & $C_n(100,200)$ & $C_n(100,500)$ & $C_n(200,100)$ \\ \hline
& \textit{1} &  0.25 (0.07,0.45) & 0.18 (0.07,0.57) & 0.17 (0.13,0.62) & 0.18 (0.10,0.45)\\
& \textit{2} &  0.22 (0.15,0.42) & 0.26 (0.19,0.45) & 0.39 (0.26,0.43) & 0.22 (0.17,0.34) \\
$\phi_{\mathbf{V}}$& \textit{3} &  0.23 (0.16,0.37) & 0.32 (0.23,0.39) & 0.39 (0.29,0.46) & 0.22 (0.14,0.35) \\
&\textit{4} &  0.20 (0.15,0.43) & 0.22 (0.16,0.38) & 0.28 (0.19,0.41) & 0.19 (0.12,0.34) \\
&\textit{5} &  0.22 (0.17,0.36) & 0.29 (0.25,0.42) & 0.37 (0.33,0.47) & 0.26 (0.17,0.33) \\ \hline
&\textit{1} &  0.03 (1.05$e^{-4}$,0.97)  &   0.01 (5.80$e^{-5}$,1.02) & 0.88 (3.62$e^{-5}$,1.07) & 0.80 (4.74$e^{-4}$,1.07) \\
&\textit{2} &  0.91 (0.31,1.17)  &  0.89 (0.45,1.17) & 0.88 (0.70,1.18) & 0.79 (0.51,1.07) \\
$\sigma_{\mathbf{V}}$&\textit{3} &  0.79 (0.46,1.14) &  0.83 (0.64,1.24) & 0.93 (0.72,1.32) & 0.89 (0.48,1.09)\\
&\textit{4} &  0.79 (1.93$e^{-3}$,1.16)  & 0.75 (0.47,1.10) & 0.81 (0.57,1.06) & 0.83 (0.35,1.15) \\
&\textit{5} &  0.85 (0.66,1.22)  & 0.96 (0.66,1.26) & 1.08 (0.84,1.28) & 0.90 (0.68,1.16)\\          \hline
&\textit{1} &  0.25 (0.20,0.32)  & 0.31 (0.22,0.34) & 0.29 (0.27,0.38) & 0.23 (0.19,0.32) \\
&\textit{2} &  0.23 (0.19,0.30)  & 0.24 (0.20,0.32) & 0.25 (0.21,0.34) & 0.22 (0.19,0.29) \\
$\phi_{\mathbf{U}}$&\textit{3} &  0.23 (0.18,0.31)  & 0.21 (0.19,0.32) & 0.25 (0.19,0.31) & 0.22 (0.18,0.31)\\
&\textit{4} &  0.24 (0.18,0.30)  & 0.25 (0.22,0.32) & 0.30 (0.24,0.35) & 0.22 (0.19,0.29) \\
&\textit{5} &  0.23 (0.18,0.32)  & 0.23 (0.19,0.30) & 0.25 (0.18,0.29) & 0.21 (0.17,0.32)\\ \hline
&\textit{1} &  0.97 (0.86,1.27)  & 1.10 (0.89,1.26) & 1.13 (1.05,1.45) & 0.96 (0.86,1.20)\\
&\textit{2} &  0.92 (0.81,1.16)  & 0.98 (0.85,1.22) & 0.99 (0.91,1.33) & 0.93 (0.83,1.18)\\
$\sigma_{\mathbf{U}}$&\textit{3} &  0.86 (0.80,1.19)  & 0.92 (0.84,1.17) & 0.99 (0.89,1.23) & 0.90 (0.82,1.18)\\
&\textit{4} &  0.95 (0.87,1.25)  & 1.00 (0.87,1.22) & 1.11 (0.97,1.40) & 0.94 (0.88,1.16)\\
&\textit{5} &  0.89 (0.78,1.11)  & 0.94 (0.81,1.09) & 0.91 (0.85,1.14) & 0.88 (0.82,1.12) \\ \hline
\end{tabular}}
\label{tab:res_sp_param}
\end{table}

\subsubsection{Evaluation of the estimation of intercept parameters}

\begin{figure}
    \centering
    \begin{subfigure}{0.6\linewidth}
        \includegraphics[width=\linewidth]{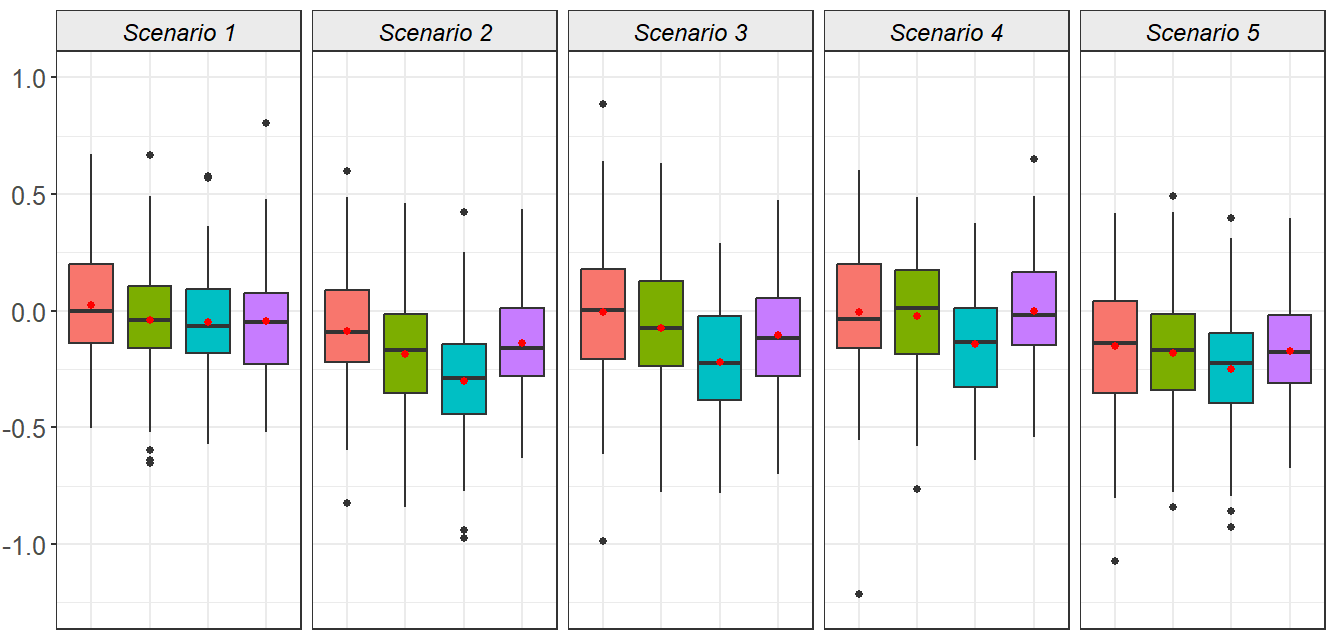}
        \caption{$\alpha'$}
        \label{fig:int_z}
    \end{subfigure}\qquad
    \begin{subfigure}{0.6\linewidth}
        \includegraphics[width=\linewidth]{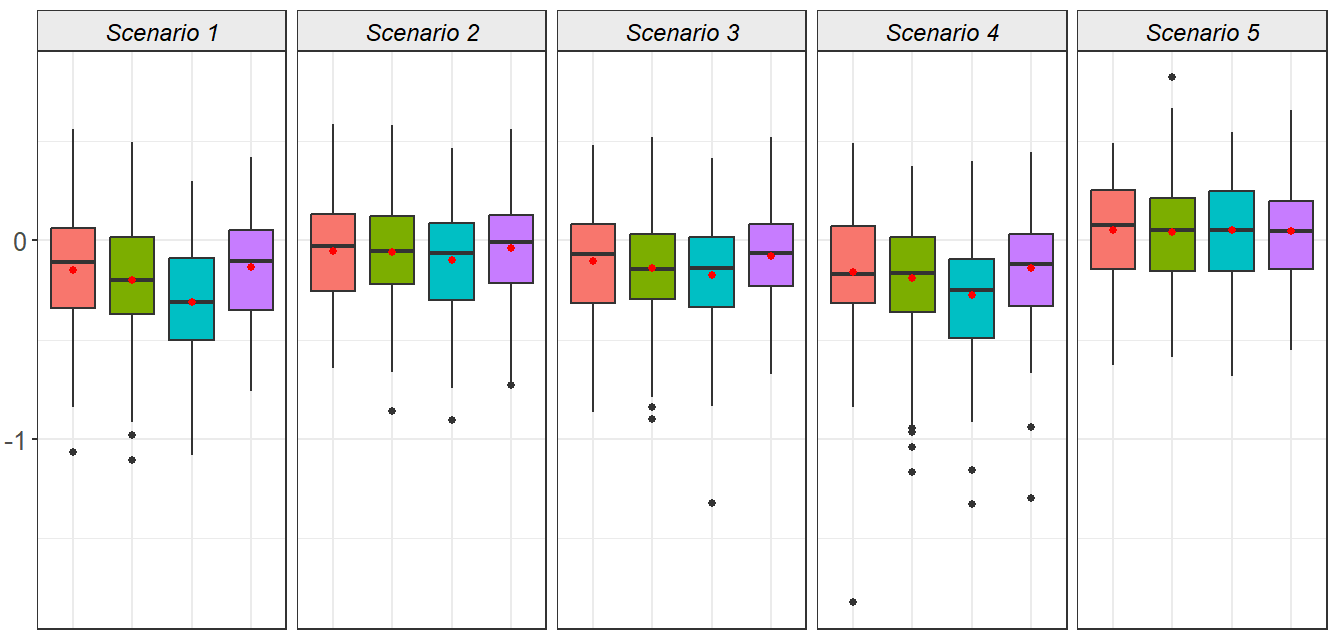}
        \caption{$\alpha$}
        \label{fig:int_y}
    \end{subfigure}
    \begin{subfigure}{0.6\linewidth}
        \includegraphics[width=\linewidth]{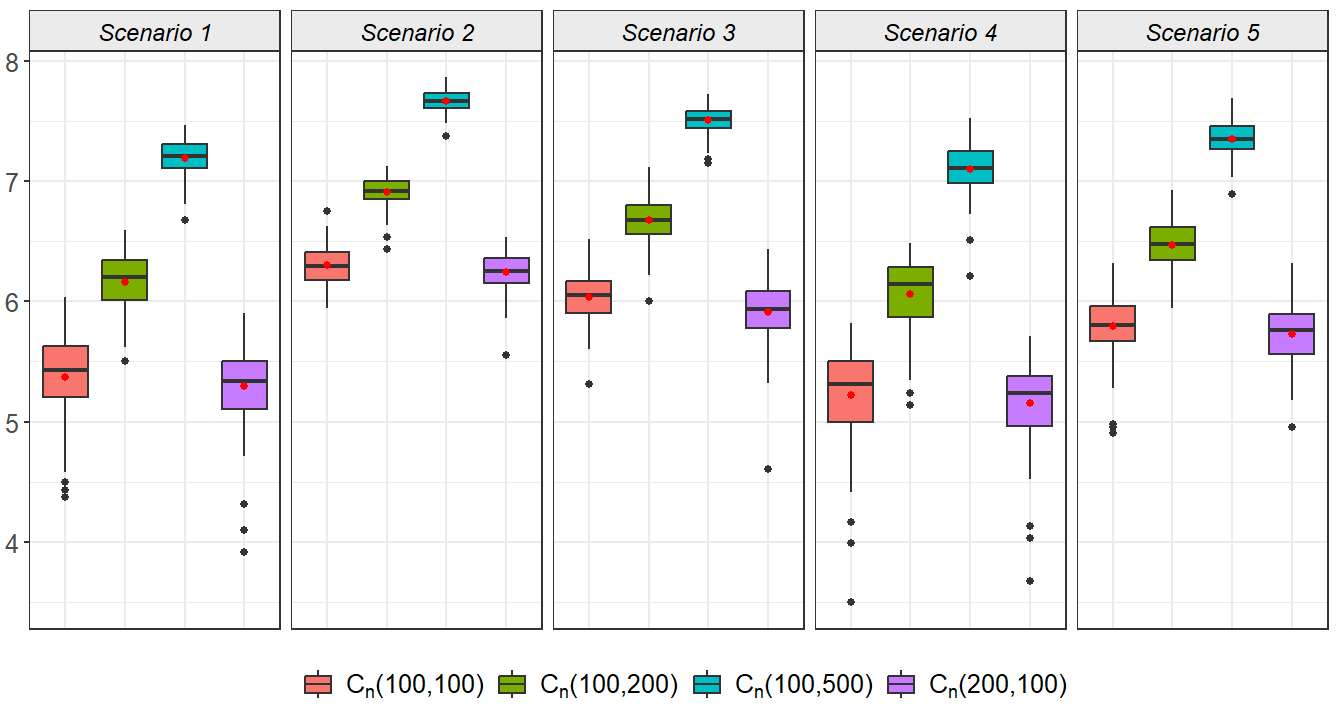}
        \caption{$\alpha''$}
        \label{fig:int_lambda}
    \end{subfigure}
    \caption{Estimates of intercept parameters ($\alpha'=0$, $\alpha=0$ and $\alpha''=0$) across sampling scenarios (\textit{Scenario 1}: $\beta'=0$ and $\beta=2$; \textit{Scenario 2}: $\beta'=1$ and $\beta=0.5$; \textit{Scenario 3}: $\beta'=1$ and $\beta=1$; \textit{Scenario 4}: $\beta'=1$ and $\beta=2$; \textit{Scenario 5}: $\beta'=2$ and $\beta=0$) and combination of samples' dimensions $C_n(n^S, n^C)$. The \textcolor{red}{red} points represent the mean values of all 100 replicas.}
    \label{fig:int_by_comb_dim}
\end{figure}

The method accurately estimates $\alpha$ and $\alpha'$, as evidenced by the inclusion of the true value (zero) within the interquartile interval across all considered model configurations (Figures \ref{fig:int_z} and \ref{fig:int_y}).
Conversely, the intercept parameter associated with the point process, denoted as $\alpha''$, exhibits a tendency toward overestimation (Figure \ref{fig:int_lambda}), becoming more biased as the dimension of the FDD increases. Referring to the definition of the intensity function \eqref{eq:ipp}, $\alpha''$ represents the expected value of the point process intensity linked to the FDD, given the zero-mean GFs $\mathbf{V}$ and $\mathbf{U}$. Since the intensity function is positive and increases with the sample size, this behavior explains why $\alpha''$ is consistently estimated above its true value (which is defined as zero) and why the bias grows as the FDD sample dimension increases.

\subsubsection{Evaluation of the prediction performance}

\begin{figure}
    \centering
    \begin{subfigure}{0.6\linewidth}
        \includegraphics[width=\linewidth]{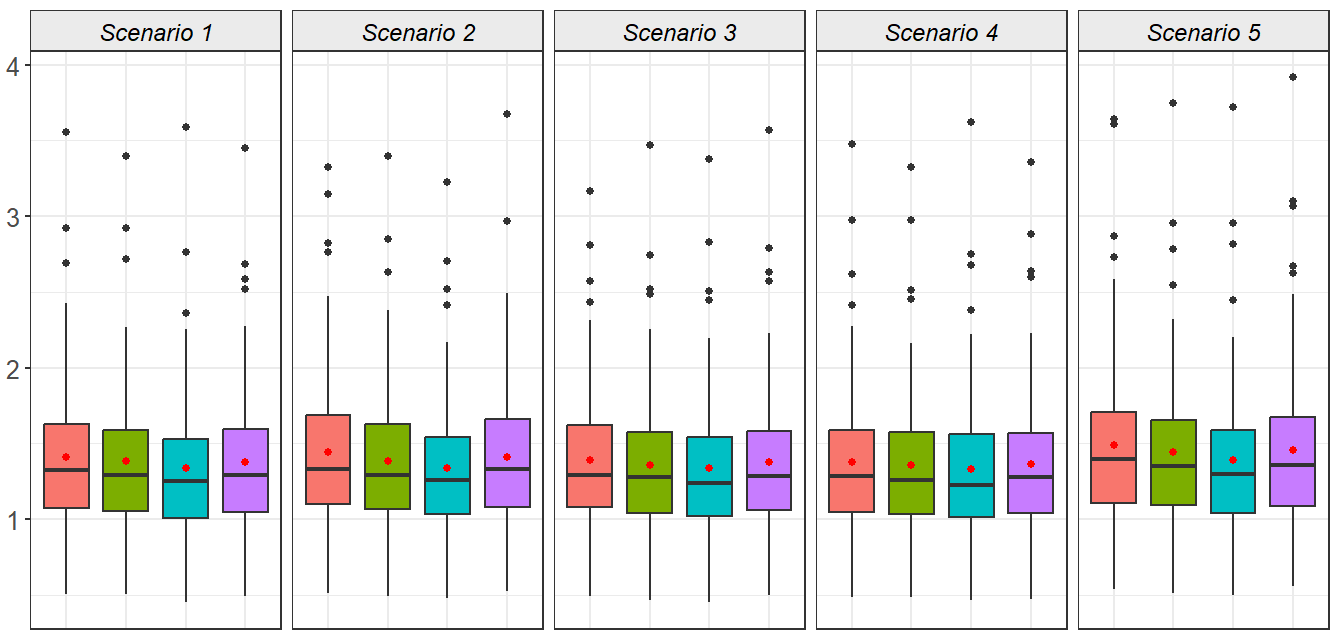}
        \caption{RMSE (Root Mean Square Error)}
        \label{fig:RMSE}
    \end{subfigure}
    \begin{subfigure}{0.6\linewidth}
        \includegraphics[width=\linewidth]{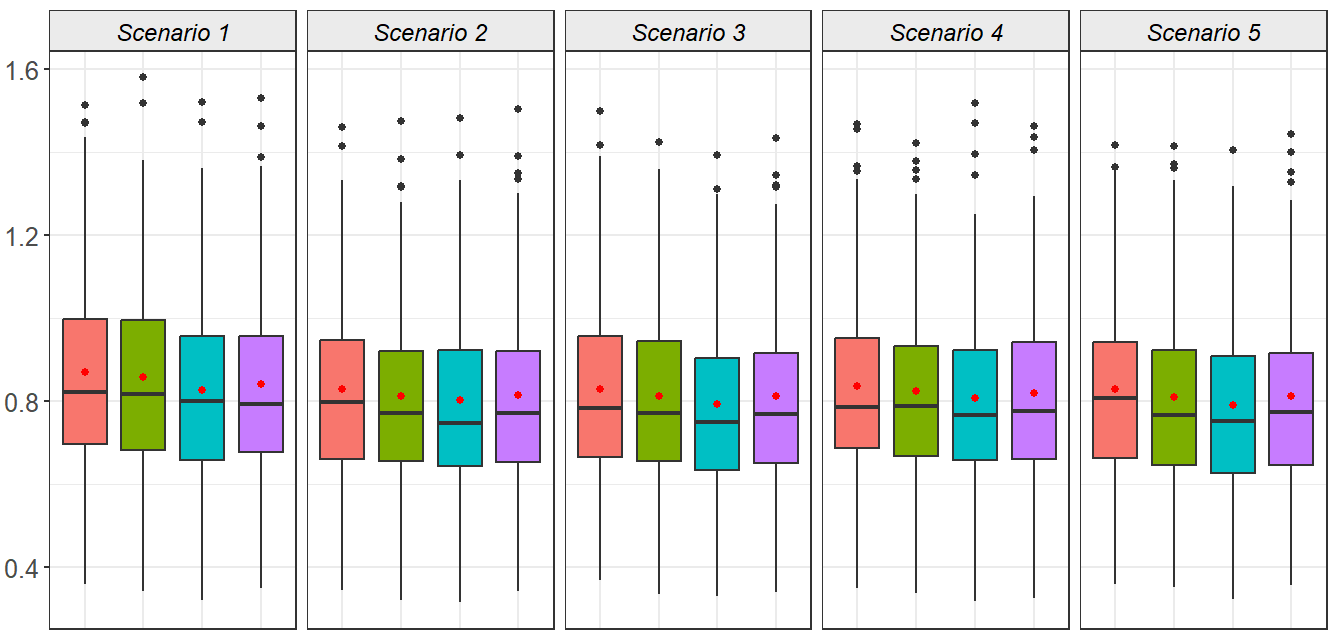}
        \caption{MAE (Mean Absolute Error)}
        \label{fig:MAE}
    \end{subfigure}
    \begin{subfigure}{0.6\linewidth}
        \includegraphics[width=\linewidth]{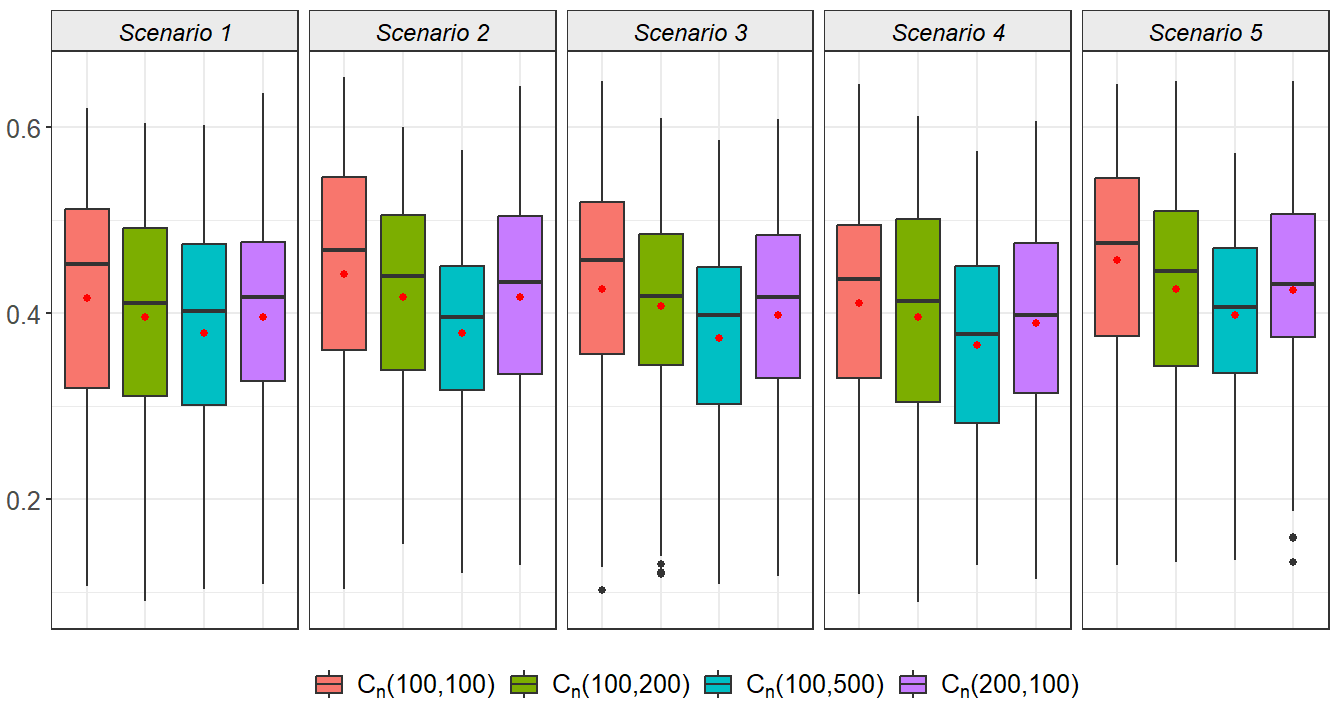}
        \caption{Hellinger distance between the observed and predicted distributions}
        \label{fig:helling}
    \end{subfigure}
    \caption{Evaluation of predictive performance. Performance metrics (RMSE, MAE, and Hellinger distance) across sampling scenarios (\textit{Scenario 1}: $\beta'=0$ and $\beta=2$; \textit{Scenario 2}: $\beta'=1$ and $\beta=0.5$; \textit{Scenario 3}: $\beta'=1$ and $\beta=1$; \textit{Scenario 4}: $\beta'=1$ and $\beta=2$; \textit{Scenario 5}: $\beta'=2$ and $\beta=0$) and combination of samples' dimensions $C_n(n^S, n^C)$. The \textcolor{red}{red} points represent the mean values of all 100 replicas.}
    \label{fig:perf_by_comb_dim}
\end{figure}

In terms of prediction performance (Figure \ref{fig:perf_by_comb_dim}), no substantial differences are discernible across various model configurations, although a modest improvement in prediction accuracy is noted with larger datasets.

Analysis of the model's predictive performance across diverse scenarios (Figure \ref{fig:helling}) reveals its capability to generate predicted data distributions closely resembling the observed ones.

\subsubsection{Evaluation of the contribution of each data source}

\begin{figure}
    \centering
    \begin{subfigure}{0.6\linewidth}
        \includegraphics[width=\linewidth]{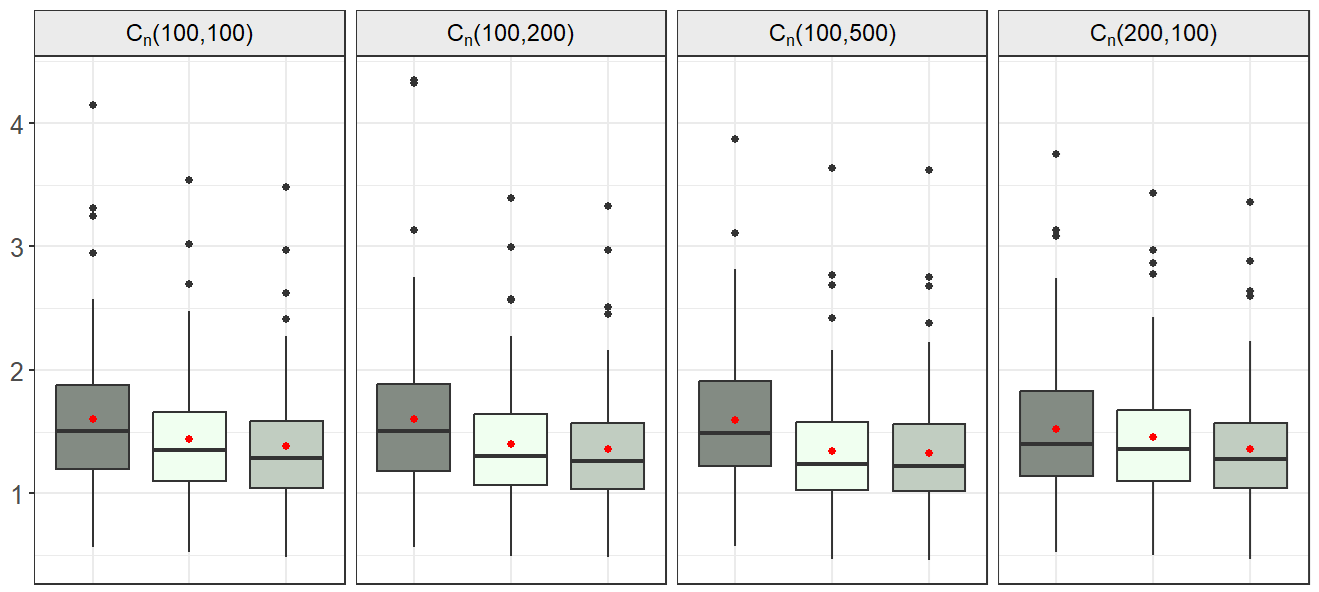}
        \caption{RMSE (Root Mean Square Error)}
        \label{fig:RMSE_3}
    \end{subfigure}
    \begin{subfigure}{0.6\linewidth}
        \includegraphics[width=\linewidth]{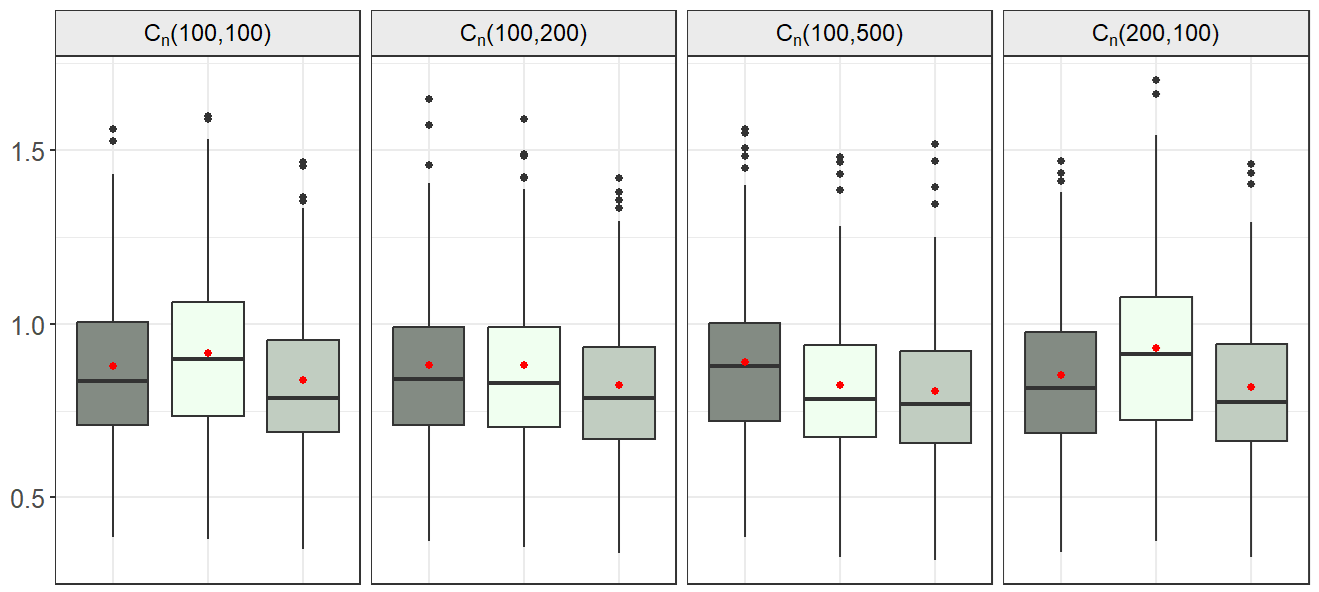}
        \caption{MAE (Mean Absolute Error)}
        \label{fig:MAE_3}
    \end{subfigure}
    \begin{subfigure}{0.6\linewidth}
        \includegraphics[width=\linewidth]{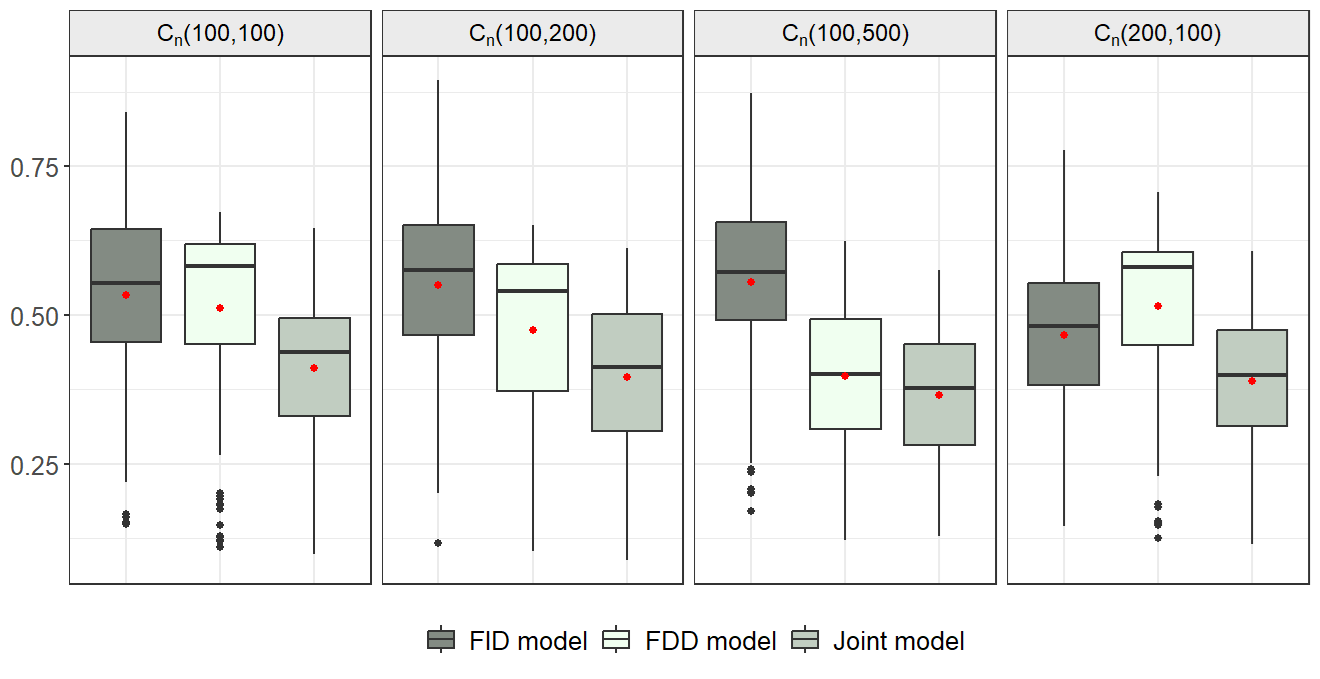}
        \caption{Hellinger distance between the observed and predicted distributions}
        \label{fig:helling_3}
    \end{subfigure}
    \caption{Evaluation of the contribution of each data source. Performance metrics (RMSE, MAE, and Hellinger distance) across model configurations and combinations of samples' dimensions  $C_n(n^S, n^C)$ in \textit{Scenario 4}. The \textcolor{red}{red} points depict the mean values of all 100 replicas.}
    \label{fig:perf_by_comb_dim3}
\end{figure}

Comparing RMSE (Figure \ref{fig:RMSE_3}) and MAE (Figure \ref{fig:MAE_3}) results, the Joint model consistently outperformed models that use one of both data sources. These discrepancy is more pronounced when analysing the results of the Hellinger distance. Larger datasets are required for accurate predictions independently of the model. Consequently, the proposed Joint model demonstrates prediction efficiency, presenting a balanced performance across various dimensions and providing robust predictions.

\begin{figure}
    \centering
    \includegraphics[width=0.8\linewidth]{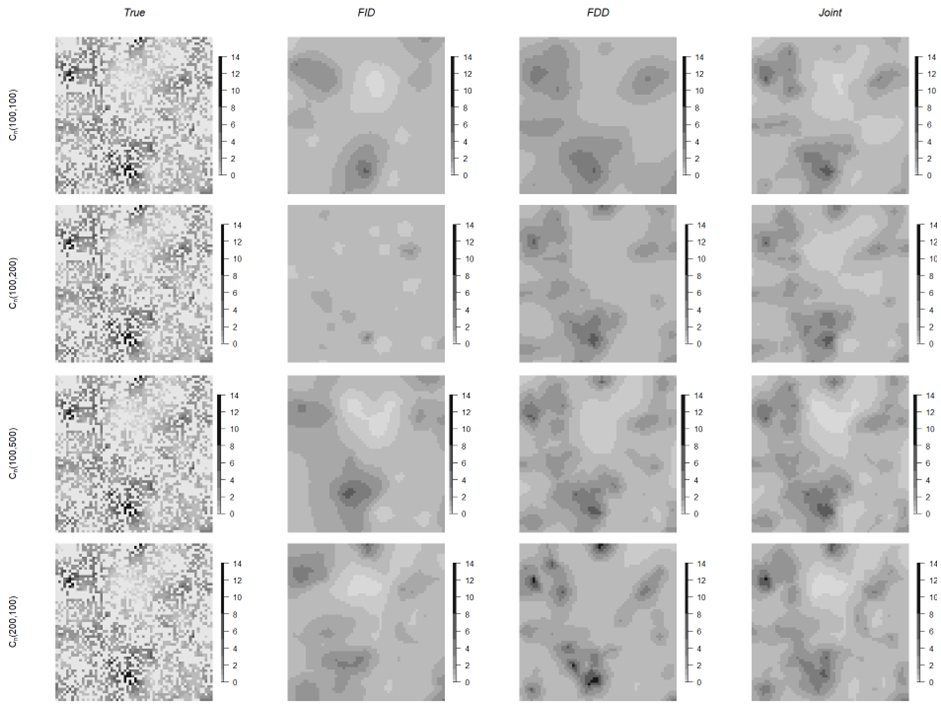}
   \caption{Results for biomass prediction based on a simulation experiment of \textit{Scenario 4} ($\beta'=1$ and $\beta=2$) across various samples' dimensions. First column: observed biomass - the true biomass values $\mathbf{S}$ observed during the simulation experiment. Second column:
FDD model prediction $\hat{\mathbf{S}}$ - results obtained from the model fitted exclusively to the FDD source. Third column: Joint model prediction $\hat{\mathbf{S}}$ - predicted biomass values resulting from modeling both FID and FDD sources. Fourth column: FID model prediction $\hat{\mathbf{S}}$ - predictions derived from the model fitted exclusively to the FID source.}
    \label{fig:sce4_by_dim}
\end{figure}

Figure \ref{fig:sce4_by_dim} illustrates the distinctions in predictions obtained from the three models (FDD, FID, and Joint) for one replica of \textit{Scenario 4}. Independently on the configuration of sample sizes, Joint model yield a predicted pattern more akin to the true one. Moreover, FDD and Joint models demonstrated enhanced capability in capturing biomass hotspots, possibly attributed to the strong dependence of the sampling process on the biomass under presence. Indeed, the influence of a strong PS dependent on the biomass process leads to higher sample values of biomass.

The predicted patterns from FDD and Joint models are quite similar, with slight variations reflecting the influence of FID. FID provides a more well-defined pattern of the response process when FID surpasses FDD in dimensionality, $C_n(200,100)$. In this specific combination, the Joint model showcases the combined contributions of both data sources, highlighting the clear impact of incorporating both FID and FDD dimensions in the Joint model for a comprehensive and nuanced representation of biomass patterns. However, it is important to note that Figure \ref{fig:sce4_by_dim} resulted from one simulated experiment and discrepancies between the models in the prediction fields can be further observed for other simulation experiments.

\subsection{Case study}

The proposed Joint model, integrating both FDD and FID, provided insights into the sardine distribution. The three models under consideration exhibited slight differences in their parameter estimates, as outlined in Table \ref{tab:res_par}, revealing the influence of each data source. Additionally, the findings shed light on the pronounced positive correlation of the sampling process from the FDD data source on both sardine presence and biomass. This is prominently illustrated in Figure \ref{fig:observ_biomass}, where the majority of FDD samples align with higher FID observations. The Joint model, in particular, produced more robust estimates for the preferential degrees, especially for $\beta$. The discrepancy in the estimation of this parameter can be attributed to the fact that the scientific survey detected sardine in areas that were not explored by the commercial fishing vessels. This suggests that incorporating FID data also contributes in defining the preferential effect associated with the FDD, as it provides additional spatial information beyond the regions covered by the fishermen.

An important disparity in parameter estimates emerges in the spatial covariance parameters $\sigma_{.}$ and $\phi_{.}$ (refer to rows 6-7 and 10-11 of Table \ref{tab:res_par}). Given that 80\% of FID observations indicate species absence, while 84\% of FDD observations are positive, an anticipated dissimilarity in variability emerges in the PAP for FID (rows 7 and 11 of Table \ref{tab:res_par}). These findings underscore the Joint model's efficacy in assimilating the variability from individual data sources, enhancing our understanding of the interplay between different datasets in the modeling process.

\begin{table}
\centering
\caption{Parameter estimates (and standard errors) involved in FID, FDD, and Joint models. Standard errors for $\sigma_.$ and $\phi_.$ are not provided since they resulted from the reparameterization of $\kappa_.$ and $\tau_.$ (see Section \ref{ssec:inf_est}).}
\begin{tabular}{lrrr}
\hline
\multicolumn{1}{c}{Parameter} & FID & FDD & Joint \\ \hline
$\alpha'$ & -10.36 (0.15) & -2.77 (0.51) & -5.52 (2.84)\\
$\alpha$ & 5.61 (0.10) & 4.81 (0.20) & 6.25 (1.04)\\
$\alpha''$ & & 1.28 (1.46) & 1.43 (1.64) \\
$\beta'$ & & 0.96 (0.03) & 0.96 (0.32) \\
$\beta$ & & 9.06 (0.03) & 0.85 (0.23) \\
$\phi_{\mathbf{V}}$ (Km) & 5.19 & 40.91 & 25.48\\
$\sigma_{\mathbf{V}}$ & 20.39 & 3.31 & 3.60\\
$log(\kappa_{\mathbf{V}})$ & -0.61 (0.61) & -2.67 (0.11) & -2.20 (0.29)\\
$log(\tau_{\mathbf{V}})$ & -3.67 (1.24) & 0.21 (0.21) & -0.35 (0.31)\\
$\phi_{\mathbf{U}}$ (Km) & 6.86 & 36.91 & 15.05\\
$\sigma_{\mathbf{U}}$ & 2.56 & 0.27 & 2.05\\
$log(\kappa_{\mathbf{U}})$ & -0.89 (0.11) & -2.57 (0.10) & -1.67 (0.29)\\
$log(\tau_{\mathbf{U}})$ & -1.32 (0.19) & 2.63 (0.18) & -0.31 (0.23)\\ \hline
\end{tabular}
\label{tab:res_par}
\end{table}

\begin{figure}
    \centering
    \includegraphics[width=0.95\textwidth]{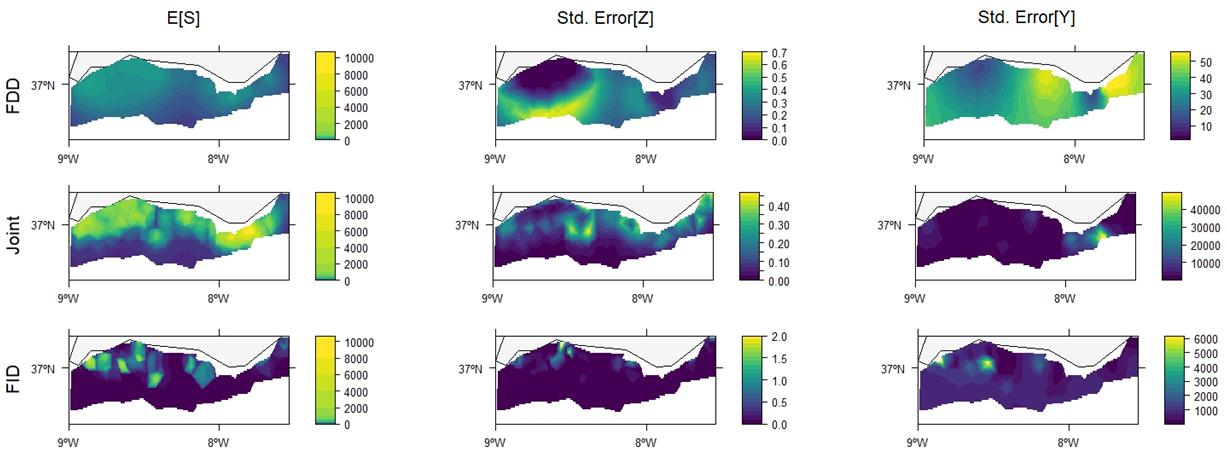}
   \caption{Predicted relative biomass index of sardine in Portuguese south coast for 2017 (first column) and standard errors associated to the predicted processes $\mathbf{Z}$ (second column) and $\mathbf{Y}$ (third column). First row: FDD model prediction - results obtained from the model fitted exclusively to the FDD (commercial) source. Second row: Joint model prediction - results from modeling both FID and FDD (survey) sources. Third row: FID model prediction - results derived from the model fitted exclusively to the FID source.}
    \label{fig:pred_biomass}
\end{figure}

The FDD and FID models yield distinct spatial prediction patterns (first column of Figure \ref{fig:pred_biomass}). The FID model reveals seven well-defined (coloured) regions that distinctly indicate species presence (first column, third row of Figure \ref{fig:pred_biomass}). In contrast, the FDD model presents a smoother pattern of relative biomass, characterized by stronger spatial dependence (first column, first row of Figure \ref{fig:pred_biomass}). Finally, predictions from the Joint model integrate features from both the FID and FDD models, underscoring the complementary contributions of each data source. Specifically, the FID data aids in delineating a clear presence-absence pattern and identifying certain hotspots, while the FDD data contributes to capturing the spatial dependency associated with both the PAP and the relative biomass process.

These distinctive contributions can be attributed to the prevalence of zero values in the FID data and a substantial number of positive values in the FDD data. Thus, the Joint model can effectively synthesize these unique contributions, offering a comprehensive spatial prediction that captures the nuances of both data sources.

\section{Discussion}

Integrating data from diverse sources to gain more comprehensive insights into spatial fish distribution poses a significant challenge in marine ecology. Both FID and FDD sources have proven their capability to offer valuable information about fish distribution \citep{Pennino2016,Izquierdo2022,Silva2023}. However, the distinct sampling processes in these datasets pose challenges when attempting joint modeling.

To address these challenges, our study introduces a two-part model that integrates FID and FDD to infer fish distribution patterns. The model is designed to accommodate ZI data and account for the unique sampling processes inherent to each dataset. We evaluated the model's strengths and limitations through a comprehensive simulation study, complemented by a case study focused on the spatial distribution of European sardine. In this context, the survey data represent FID, while commercial catch records represent FDD. These two datasets exhibit discrepancies; for example, 80\% of FID observations indicate sardine absence, whereas 84\% of FDD observations show sardine presence. This disparity arises primarily from the underlying sampling processes, as fishermen typically rely on prior knowledge and real-time observation of fish distribution, leading to a higher proportion of presence observations in FDD. Such discrepancies further highlight the necessity of a joint modeling approach to capture a more comprehensive picture of sardine distribution.

Our proposed model exhibits a strong capacity to produce accurate estimates for preferential parameters, attesting to its reliability across various scenarios. However, the model presents slight difficulty in estimation in scenarios with strong PS effect on PAP and biomass process, as also observed by \citet{Silva2024}, highlighting the impact of preferential factors on parameter estimation. Examining spatial covariance estimation reveals a consistent pattern of accurate estimation, except the overestimation of $\phi_{\mathbf{V}}$. Additionally, the mode consistently exhibit accuracy across all model configurations for estimating the intercept parameters, excepting the intercept parameter $\alpha''$ for the intensity function that tends to be overestimated due to its role of representing the intensity mean of the point process.

Prediction performance consistently favors models that incorporate both FID and FDD over those utilizing a single data source across all examined scenarios, demonstrating the model’s reliability in accurately capturing underlying distribution patterns. When FID and FDD sources share identical dimensions, the single source both presents a very little difference in their performance, suggesting that preferential and non-preferential sampling can perform equally well as long as the possibility of PS is considered. Results from both the simulation study and the empirical application indicate that the FDD and Joint models present a higher capability to capture the spatial correlation of the response process, while FID model may bring information about marine regions not explored by fishermen. Thus, the Joint model integrates contributions from both data sources, providing a more comprehensive representation of biomass patterns.

Although the model introduced by \cite{Rufener2021} demonstrated satisfactory results in predicting fish distribution, it is important to note that the authors did not account for the PS nature of commercial data. However, it is widely recognized that a substantial portion of opportunistic data, such as derived from commercial sources, exhibits relevant PS. Ignoring PS in spatial prediction may introduce significant bias, as emphasized by \cite{Diggle2010}. A limitation of this model arises in its suitability for handling ZI data since it is tailored for count data, using the Negative Binomial distribution to accommodate abundance data.

\cite{Alglave2022} demonstrated the robustness and consistency of their proposed model across a spectrum of scenarios, effectively addressing the ZI issue in the data. However, an important aspect not explicitly considered in their model is the potential variability in conditions governing PAP and biomass process, as observed in \citet{Silva2023}. In scenarios where local fishing intensity is contingent on fish presence and fish biomass in distinct ways, the model assumption of a unique relationship between the sampling process and the relative biomass field may not capture such variations. In our study, we address this limitation by proposing a two-part model that allows for the differentiation between PAP and biomass process. This consideration enhances the comprehensiveness of our model and contributes to a more detailed understanding of the complex spatial dynamics governing fish distribution in marine environments.

The significance of environmental conditions in influencing spatial species distribution is well-established \citep{AUSTIN20071}. Incorporating explanatory covariates representing these conditions in our model is not only important for achieving more precise predictions but also offers valuable insights into the complex relationship between the species and the ecosystem \citep{Hefley2016}. As such, a key avenue for future development involves the inclusion of covariates in the model since this consideration is crucial as both the PAP and biomass process can be elucidated by a set of environmental conditions \citep{Pennino2020,Silva2023}. Moreover, the integration of additional environmental and external variables may be relevant to comprehensively describe the preferentiality for certain locations \citep{Manceur2014,Pennino2019}. Factors such as distance to the coast and bathymetry can significantly influence the spatial arrangement of fishing locations, since the fishermen tend to stay closer to the port/coast as possible due to fuel costs, contributing to a more detailed understanding of the preferential sampling dynamics. This enhancement will not only refine the precision of our predictions but also contribute to a more comprehensive and ecologically informed model. 

Beyond the spatial dimension, the temporal scale is a critical component in species distribution modeling, as the temporal evolution holds significant ecological relevance \citep{Paradinas2017,Minaya2018}. Consequently, another point for future investigation involves extending our proposed model to a spatio-temporal framework, enabling the prediction of temporal trends, seasonal variations, and long-term ecological patterns that are integral to a thorough understanding of species distribution dynamics.

In terms of practical biological applications, a potential future direction is to separately model juvenile and adult abundance indexes. Juvenile fish might be avoided by fishermen due to low fishing value and restrictions, and these areas likely correspond to higher biomass, as indicated in the FID, which could impact parameter estimates.

\appendix
\counterwithin{equation}{section}
\counterwithin{figure}{section}
\section{Likelihood of the model}
\label{sec:theory}

The joint distribution \eqref{eq:dist} is determined by the distribution of the biomass process under the presence (conditioned on the GMRF $\mathbf{U}$ and the sampling processes $\mathbf{X}^S$ and $\mathbf{X}^C$), the distribution of the PAP (conditioned on the GMRF $\mathbf{V}$ and the sampling processes $\mathbf{X}^S$ and $\mathbf{X}^C$), the distributions of the sampling processes $\mathbf{X}^C$ (conditioned on the GMRFs $\mathbf{U}$ and $\mathbf{V}$) and $\mathbf{X}^S$, and the distributions of both GFs $\mathbf{U}$ and $\mathbf{V}$. Each of these distributions is characterized by a specific expression, with the distribution of $\mathbf{X}^S$ remaining constant as it is assumed as a HPP.
\begin{align}
\left[\mathbf{Y},\mathbf{Z},\mathbf{X}^{C},\mathbf{X}^{S},\mathbf{U},\mathbf{V} \right]  = &
\left [\mathbf{Y} \vert \mathbf{X}^{C},\mathbf{X}^{S},\mathbf{U} \right ]  \left [\mathbf{Z} \vert \mathbf{X}^{C},\mathbf{X}^{S},\mathbf{V} \right ] \left [\mathbf{X}^{C},\mathbf{X}^{S},\mathbf{U},\mathbf{V} \right]\nonumber\\ = &\left [\mathbf{Y} \vert \mathbf{X}^{C},\mathbf{X}^{S},\mathbf{U} \right ]  \left [\mathbf{Z} \vert \mathbf{X}^{C},\mathbf{X}^{S},\mathbf{V} \right ] \left [ \mathbf{X}^{C} \vert \mathbf{U}, \mathbf{V} \right] \left [\mathbf{U}\right ]
\left [\mathbf{V}\right ] \left [\mathbf{X}^{S}\right ].
\label{eq:dist} 
\end{align}

Given the result presented in \eqref{eq:dist} and denoting the space of parameters as $\Theta$, the likelihood of the model is
\begin{align}
\mathcal{L}(\Theta)=&\mathcal{L}(\mu, \sigma; \mathbf{y}) \times \mathcal{L}(\pi; \mathbf{z}) \times \mathcal{L}(\lambda; \mathbf{x}) \times  \mathcal{L}(\sigma_U, \phi_U) \times \mathcal{L}(\sigma_V, \phi_V),
\label{eq:lik}
\end{align}
where $\mathcal{L}(\mu, \sigma; \mathbf{y})$ represents the likelihood for $\mathbf{Y} \vert \mathbf{X}^{C},~\mathbf{X}^{S},\mathbf{U}$ \eqref{eq:lik_y}, the likelihood for $\mathbf{Z} \vert \mathbf{X}^{C},\mathbf{X}^{S},\mathbf{V}$ is denoted by $\mathcal{L}(\pi; \mathbf{z})$ \eqref{eq:lik_z}, $\mathcal{L}(\lambda; \mathbf{x})$ identifies the likelihood for $\mathbf{X}^{C} \vert \mathbf{U}, \mathbf{V}$ \eqref{eq:lik_lambda}, and  $\mathcal{L}(\sigma_U, \phi_U)$ and $\mathcal{L}(\sigma_V, \phi_V)$ represent the likelihoods for $\mathbf{U}$ and $\mathbf{V}$ (\eqref{eq:lik_u} and \eqref{eq:lik_v}), respectively. Hence, the joint log-likelihood is given by
\begin{align}
\ell(\Theta)=&\ell(\mu, \sigma; \mathbf{y}) + \ell(\pi; \mathbf{z})+ \ell(\lambda; \mathbf{x}) + \ell(\sigma_U, \phi_U) +\ell(\sigma_V, \phi_V).
\end{align}

\subsection{Likelihood for $\mathbf{Y} \vert \mathbf{X}^{C},\mathbf{X}^{S},\mathbf{U}$}
\begin{align}
\mathcal{L}(\mu, \sigma; \mathbf{y}) = & \prod_{i=1}^{n} \frac{y_i^{\left( \frac{\mu^2}{\sigma^2}-1 \right)} e^{-\left(\frac{\mu}{\sigma^2}\right)y_i} \left( \frac{\mu}{\sigma^2} \right)^{\frac{\mu^2}{\sigma^2}} }{\Gamma(\frac{\mu^2}{\sigma^2})}.
\label{eq:lik_y}
\end{align}

The corresponding log-likelihood is given by
\begin{align}
\ell(\mu,\sigma; \mathbf{y}) & = \sum_{i=1}^n \left( \left(\frac{\mu^2}{\sigma^2}-1\right)log(y_i)-\frac{\mu}{\sigma^2} y_i + \frac{\mu^2}{\sigma^2} log\left(\frac{\mu}{\sigma^2}\right) - log \left( \Gamma\left(\frac{\mu^2}{\sigma^2}\right) \right) \right) \nonumber \\
& = n\frac{\mu^2}{\sigma^2} log\left(\frac{\mu}{\sigma^2}\right)-nlog\left(\Gamma\left(\frac{\mu^2}{\sigma^2}\right)\right)+\sum_{i=1}^{n} \left( \left(\frac{\mu^2}{\sigma^2}-1\right)log(y_i)-\frac{\mu}{\sigma^2} y_i \right) 
\end{align}
where $n$ represents the dimension of all data ($n=n^S+n^C$).

\subsection{Likelihood for $\mathbf{Z} \vert \mathbf{X}^{C},\mathbf{X}^{S},\mathbf{V}$}

\begin{align}
\mathcal{L}(\pi; \mathbf{z}) = \prod_{i=1}^{n} \pi^{z_i}(1-\pi)^{1-z_i}
= \pi^{\sum_{i=1}^{n}z_i}(1-\pi)^{n-\sum_{i=1}^{n}z_i}
\label{eq:lik_z}
\end{align}

and, hence, the log-likelihood is given by
\begin{align}
\ell(\pi; \mathbf{z})  = log(\pi)\sum_{i=1}^n z_i + log(1-\pi) \left(  n-\sum_{i=1}^n z_i \right).
\end{align}

\subsection{Likelihood for $\mathbf{X}^{C} \vert \mathbf{U}, \mathbf{V} $}$~$\\

Following \cite{Diggle2013}, the likelihood for IPP comes from
\begin{align}
\mathcal{L}(\lambda; \mathbf{x}) = & \prod_{i=1}^{n^C} \frac{e^{-\omega}\omega^{n^C}}{n^C!}\times \frac{\lambda(\mathbf{x}_i^C)}{\omega_t}.
\label{eq:lik_lambda}
\end{align}

The log-likelihood is expressed as
\begin{align}
\ell(\lambda; \mathbf{x})  = & ~log \left( \frac{e^{-\omega}\omega^{n^C}}{n^C!} \right)+ \sum_{i=1}^{n^C}
log \left( \frac{\lambda(\mathbf{x}_i)}{\omega} \right)\\ = & -\omega + n^C \times log(\omega)-log(n^C!)+\sum_{i=1}^{n^C} \left( log(\lambda(\mathbf{x}_i))-log(\omega) \right)\nonumber \\ \simeq  & \sum_{i=1}^{n^C} log(\lambda(\mathbf{x}_i))-\omega = \sum_{i=1}^{n^C} log(\lambda(\mathbf{x}_i))-\int_{\mathcal{A}} \lambda(\mathbf{s})\partial \mathbf{s}.
\end{align}

\subsection{Likelihoods for $\mathbf{U}$ and $\mathbf{V}$}
\begin{align}
\mathcal{L}(\sigma_U, \phi_U) = & \frac{1}{(\sqrt{2\pi})^N \vert \Sigma_U \vert ^{1/2}}exp \lbrace -\frac{1}{2}\mathbf{u}_.^{\texttt{T}} \Sigma_U^{-1} \mathbf{u}_. \rbrace,
\label{eq:lik_u}
\end{align}
\begin{align}
\mathcal{L}(\sigma_V, \phi_V) = & \frac{1}{(\sqrt{2\pi})^N \vert \Sigma_V \vert ^{1/2}}exp \lbrace -\frac{1}{2}\mathbf{v}_.^{\texttt{T}} \Sigma_V^{-1} \mathbf{v}_. \rbrace.
\label{eq:lik_v}
\end{align}

and, hence, the corresponding log-likelihoods are given by
\begin{align}
\ell(\sigma_U, \phi_U)  = & -\frac{N}{2}log(\pi) -\frac{log(\vert \Sigma_U \vert)}{2} - \frac{1}{2}\mathbf{u}_.^{\texttt{T}} \Sigma_U^{-1} \mathbf{u}_.,
\end{align}
\begin{align}
\ell(\sigma_V, \phi_V)  = & -\frac{N}{2}log(\pi) -\frac{log(\vert \Sigma_V \vert)}{2} - \frac{1}{2}\mathbf{v}_.^{\texttt{T}} \Sigma_V^{-1} \mathbf{v}_.
\end{align}
where $N$ denotes the dimension of the prediction grid (or mesh).\\

\section{Impact of sample dimension on observation process}
\label{app:B}

\begin{figure}[h]
    \centering
    \includegraphics[width=0.7\textwidth]{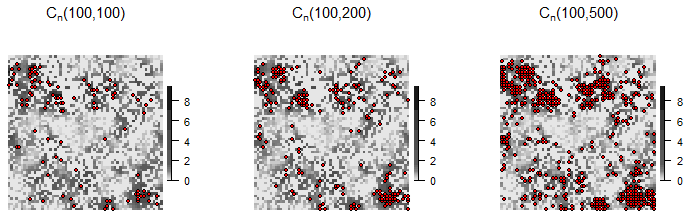}
    \caption{Examples of simulated intensity fields and FDD locations for Scenario 4 across combinations of samples' dimensions $C_n(100,100)$, $C_n(100,200)$ and $C_n(100,500)$.} 
    \label{fig:sample_by_dim}
\end{figure}

\textbf{Acknowledgments:} The authors would like thanks the teams that collected the data. The biological survey datasets generated during and/or analyzed during the current study are available from the Portuguese Data Collection Framework on reasonable request. It is not ethically feasible to share any AIS data, as it would publicly reveal vessel information indicating where the activity takes place, while disclosing sensitive information. Additionally, the AIS data outputs are ruled by a confidentiality agreement between the different authors, preventing the share of the provided AIS data outputs, any private information regarding the fishing vessel and other related information.\\

\textbf{Funding:} This study received support from Portuguese funding provided through the Centre for Mathematics via the following projects: DOI {10.54499/UIDP/00013/2020}, DOI {10.54499 /UIDB/00013/2020}, and the Portuguese Foundation for Science and Technology (FCT) through the Individual PhD Scholarship PD/BD/150535/2019, the research grant UIDB/04292/2020, and the project PTDC/MAT-STA/28243/2017. Additionally, support was provided by the {SAR-DINHA2030} project (MAR-111.4.1-FEAMPA-00001).

\bibliographystyle{apalike}
\bibliography{Manuscript}

\end{document}